\documentclass[aps,pre,preprint,showpacs,showkeys,floatfix,superscriptaddress,]{revtex4-1}

\usepackage{amsmath}
\usepackage{amssymb}
\usepackage{nccmath}
\usepackage[none]{hyphenat}
\usepackage[utf8]{inputenc}
\usepackage{graphicx}
\usepackage[caption=false]{subfig}
\usepackage[export]{adjustbox}
\usepackage[percent]{overpic}
\usepackage{natbib}
\usepackage[dvipsnames]{xcolor}
\newcommand{\valmed}[1]{\langle #1 \rangle}

\begin{document}

\title{Recovery of contour nodes in interdependent directed networks}

\author{Ignacio A. Perez}
\email{Correspondence: ignacioperez@mdp.edu.ar}
\affiliation{Instituto de Investigaciones F\'isicas de Mar del Plata
(IFIMAR)-Departamento de F\'isica, FCEyN, Universidad Nacional de
Mar del Plata-CONICET, De\'an Funes 3350, (7600) Mar del Plata, Argentina}
\author{Cristian E. La Rocca}
\affiliation{Instituto de Investigaciones F\'isicas de Mar del Plata
(IFIMAR)-Departamento de F\'isica, FCEyN, Universidad Nacional de
Mar del Plata-CONICET, De\'an Funes 3350, (7600) Mar del Plata, Argentina}

\begin{abstract}
    \noindent

    Extensive research has focused on studying the robustness of interdependent non-directed networks and
    the design of mitigation strategies aimed at reducing disruptions caused by cascading failures. However,
    real systems such as power and communication networks are directed, which underscores the necessity of
    broadening the analysis by including directed networks. In this work, we develop an analytical framework 
    to study a recovery strategy in two interdependent directed networks in which a fraction $q$ of nodes
    in each network have single dependencies with nodes in the other network. Following the random failure
    of nodes that leaves a fraction $p$ intact, we repair a fraction of nodes that are neighbors of the giant
    strongly connected component of each network with probability or recovery success rate $\gamma$. Our 
    analysis reveals an abrupt transition between total system collapse and complete recovery as $p$ is 
    increased. As a consequence, we identify three distinct phases in the $(p, \gamma)$ parameter space: 
    collapse despite intervention, recovery enabled by the strategy, and resilience without intervention. 
    Moreover, we demonstrate our strategy on a system built from empirical data and find that it can save
    resources compared to a random recovery strategy. Our findings underscore the potential of targeted 
    recovery strategies to enhance the robustness of real interdependent directed networks against cascading
    failures.
    
\end{abstract}

\maketitle

\section{Introduction}

Most of real-world systems usually interact with one another through dependencies. For instance, critical
infrastructures such as power grids and water supply systems depend on communication networks for operational
control, while communication networks require electricity to function and water systems are used to cool power
generators. Dependencies can create feedback loops that are not captured by traditional single-network models. 
Thus, interdependent networks, which have connectivity links within each single network and dependency links
between different networks, exhibit structural and dynamical behaviors distinct from isolated 
systems~\cite{rin-01,ros-08,ves-10,bas-12,bru-12}. While interdependencies can enhance the overall performance
of networks, they may also introduce vulnerabilities. A small system malfunction can propagate back and forth
between two (or more) interdependent systems in a process known as cascading 
failures~\cite{boc-14,kiv-14,dag-14,val-20}. Incidents such as the faulty software update from 
CrowdStrike~\cite{dut-24} in July 2024, which led to widespread disruptions of airlines, banks, broadcasters,
healthcare providers, etc., hurricane Katrina in USA~\cite{lea-06}, where oil rigs and refineries were 
destroyed, causing an exponential rise in the price of fuel, or the massive 2003 blackout in 
Italy~\cite{sfo-06}, which affected transportation and communications networks, underscore the need for 
studying collapse in interdependent networks in order to build more robust systems and design effective response
strategies. For example, the restoration of electric power systems is particularly critical, as these systems
are essential for the operation and management of nearly all other infrastructures. In practice, certain 
electric transmission substations are often prioritized during recovery efforts because they support critical 
facilities, such as airports, hospitals, and emergency services. This highlights the importance of effective 
recovery strategies in minimizing the impact of cascading failures and ensuring the resilience of 
interconnected systems.

In a pioneering work, Buldyrev et. al.~\cite{bul-10} studied cascading failures in two fully interdependent
non-directed networks by developing an analytical framework based in node percolation~\cite{sta-94,cal-00}, a
process extensively used for modeling propagation phenomena. In isolated networks, random node percolation 
occurs when nodes are removed at random leaving a fraction $p$ intact, producing the network breakdown into 
clusters of connected nodes with different size. In this process, the giant component (GC), i.e. the cluster of 
biggest size $P_{\infty}$, undergoes a continuous transition at a critical point $p_c$, which separates a 
collapsing phase ($P_{\infty} = 0$ for $p \leq p_c$) from a non-collapsing phase ($P_{\infty} > 0$ for 
$p > p_c$). Buldyrev et. al. found that interdependent systems have larger values of $p_c$ and thus are more
fragile than isolated networks, when the degree distribution is the same. Moreover, they found that 
interdependent networks present abrupt first-order transitions, in which $P_{\infty}$ changes from a finite 
value to zero with a small variation of the fraction $p$ of initial remaining nodes. 

Several modifications have been made to the original model~\cite{bul-10} for non-directed networks to study 
system robustness~\cite{par-10,par-11,gao-11,hua-11,don-13,sch-13,val-14,don-19,don-21} or with the focus on 
designing mitigation and recovery strategies~\cite{gon-15,hu-16,lar-18}, among other goals. For instance, Di 
Muro et. al.~\cite{dim-15} studied two interdependent networks where failed nodes that are neighbors of the GC 
of each network are repaired with probability $\gamma$. This is a reasonable strategy given the existing
facilities to do so in many real systems such as transportation networks, where it is easier to bring the
necessary equipment for repairing a damaged site through the transportation system itself. Their results show
abrupt transitions between complete collapse ($P_{\infty} = 0$) and full functionality ($P_{\infty} = 1$).
More precisely, they found three distinct phases: a phase in which the system never collapses without being
restored, another phase where breakdown is avoided due to the strategy, and a phase in which system collapse
cannot be avoided, even if the strategy is implemented.

While significant progress has been achieved in the study of non-directed systems, research efforts have only
recently shifted toward exploring interdependent directed networks. Many real-world systems are inherently 
directed, such as transportation networks and power grids~\cite{bie-10,bie-11}, biological 
networks~\cite{coh-90,wat-98,fel-00,jon-02}, and the World Wide Web~\cite{bro-00}, and it is known that 
directed links (e.g., from generators to substations that lower energy voltage in a power grid) alter network
structure and can have a significant effect on failure propagation~\cite{new-01,dor-01,schw-02}. Thus, 
research on interdependent directed networks has initially focused on studying the robustness of systems under
different types of failures or topologies~\cite{liu-16,liu-19,xu-21,lv-23}. We believe that, in addition, it 
is necessary to explore mitigation or recovery strategies that can help to develop more resilient 
interdependent directed systems against cascading failures.

In this work, we model a process of cascading failures in an interdependent system composed of two directed
networks, where a fraction $q$ of nodes in each network have single dependencies with nodes from the other 
network. Inspired by the work in~\cite{dim-15}, we implement a recovery strategy where a subset of the nodes
belonging to the contour of the giant strongly connected component (GSCC) of each network are repaired with
probability $\gamma$. This probability can be interpreted as the technical ability to repair nodes or the rate
of success, in the best-case scenario where resources are always available. Using random node percolation and
generating functions, we develop an analytical approach to compute the size of the mutual giant strongly 
connected component (MGSCC), $P_{\infty}$, at the end of the process, after an initial random failure of a 
fraction $1 - p$ of nodes in one of the networks. Moreover, we build phase diagrams for $P_{\infty}$ on the 
plane $(p, \gamma)$, which allow us to recognize different system behaviors. In addition, we simulate the 
process on a system built from empirical data and contrast the proposed strategy with the random recovery of 
nodes, in order to demonstrate the practical application of our model in a realistic scenario and the 
advantages with respect to other known recovery strategies.

\section{Model}

\begin{figure}
    \centering
    \includegraphics[width=0.5\linewidth]{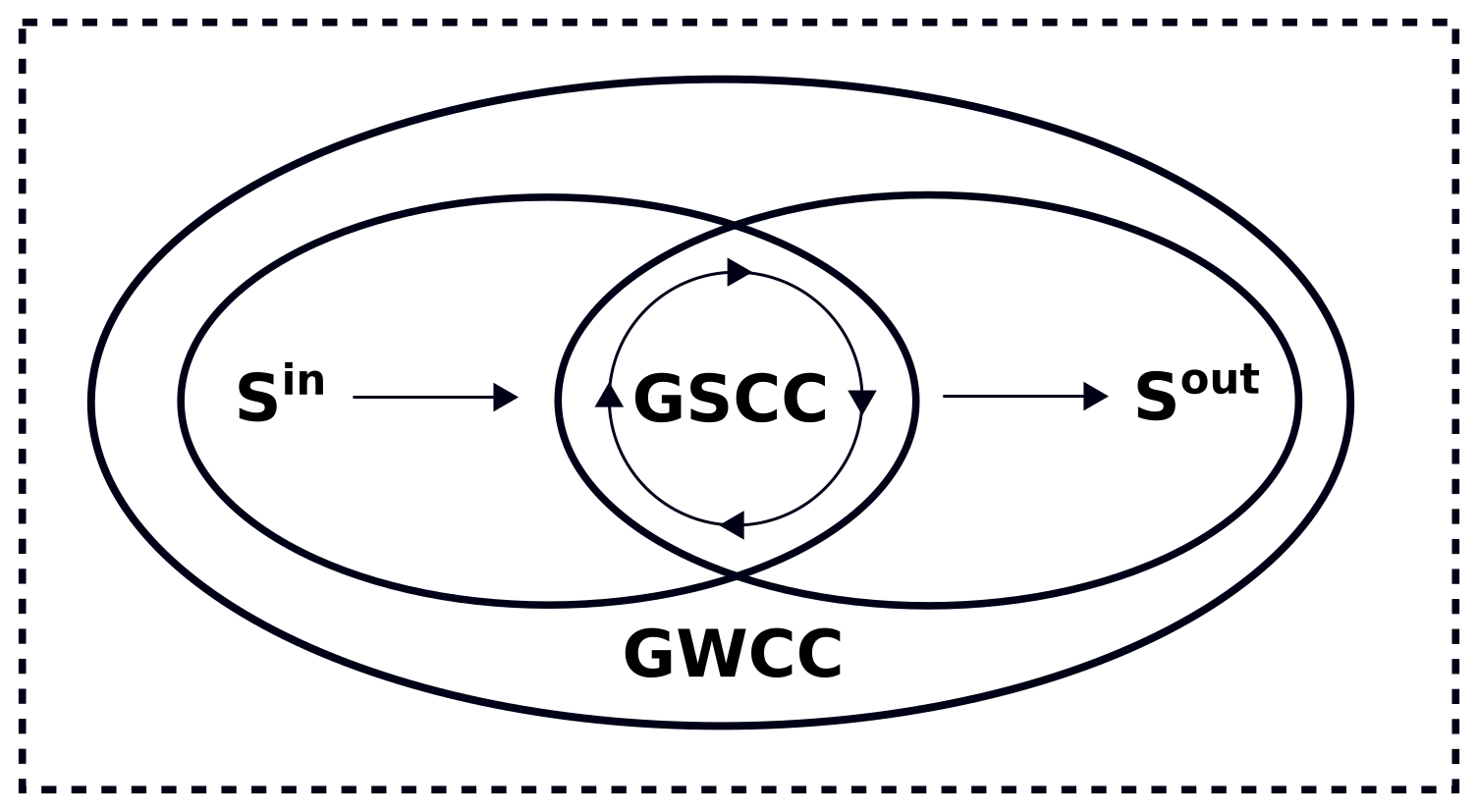}
    \caption{Scheme of the structure of directed networks. A giant weakly connected component (GWCC) arises,
    where nodes can reach other nodes through paths that ignore directions. Within this component, a smaller
    but strongly connected component (GSCC) contains nodes that can follow a directed path to every other 
    node in the component. Thus, closed paths of directed links form between each pair of nodes in the GSCC.
    Then, nodes that can arrive to the GSCC through directed paths belong to the S\textsuperscript{in}
    component, while nodes reached from the GSCC via directed paths belong to the S\textsuperscript{out}
    component.}
    \label{dir-net-comps}
\end{figure}

Our system consists in two interdependent directed networks A and B with the same size $N$. Within each 
network or layer, nodes are connected via directed connectivity links, where $P^{\ in}_i(k^{\ in}_i)$ and
$P^{\ out}_i(k^{\ out}_i)$, $i = A, B$, are the uncorrelated degree distributions for incoming and outgoing
links and $k_{in/out}$ represents the number of incoming/outgoing links that a node can have in network $i$.
To build a given layer, we use a slight variation of the configuration model~\cite{mol-95,new-01,bog-05} with
the condition $\valmed{k_{in}} = \valmed{k_{out}}$ (see Appendix~\ref{app:dir-net}). The structure of a 
directed network can be mainly described by the following three components that add to the giant weakly 
connected component (GWCC), the largest cluster where each node can reach any other node in the cluster by 
following one path ignoring link directions~\cite{dor-01}:
\begin{enumerate}
    \item Giant strongly connected component (GSCC): The largest component of connected nodes where each
    node can reach any other node in the cluster by following one directed path. In the process of
    cascading failures, we regard the GSCC as the functional component within each network. 
    \item Giant in-component (S\textsuperscript{in}): Contains nodes that can reach the GSCC by following
    a directed path.
    \item Giant out-component (S\textsuperscript{out}): Contains nodes that can be reached from the GSCC
    by following a directed path.
\end{enumerate}
By definition, the GSCC is contained in both S\textsuperscript{in} and S\textsuperscript{out}, 
GSCC = S\textsuperscript{in} $\cap$ S\textsuperscript{out}. These components are illustrated in 
Fig.~\ref{dir-net-comps}.
 
Once connectivity links within each network are set, we build the interdependencies between networks by
imposing a fraction $q_A$ of nodes in network A to depend on nodes from network B and a fraction $q_B$ of
B-nodes to depend on A-nodes, in the following manner:
\begin{enumerate}
    \item A-nodes (B-nodes) with dependencies depend only on a single, randomly selected B-node (A-node).
    \item If different A-nodes (B-nodes) have dependencies, these dependencies correspond to different 
    B-nodes (A-nodes).
    \item If node $i$ from A (B) depends on node $j$ from B (A), and if node $j$ depends on node $k$ from
    A (B), then $i = k$ (known as the no-feedback condition~\cite{bul-10,gao-11-bis}).
\end{enumerate}
In this way, dependencies can be unidirectional or bidirectional, the latter of which are allowed by
condition 3. It is important to note, as we state in Appendix~\ref{app:dir-net}, that for the internal
connectivity links within each network there are no bidirectional connections.

\begin{figure}
    \centering
    \subfloat{\begin{overpic}[width=\linewidth]{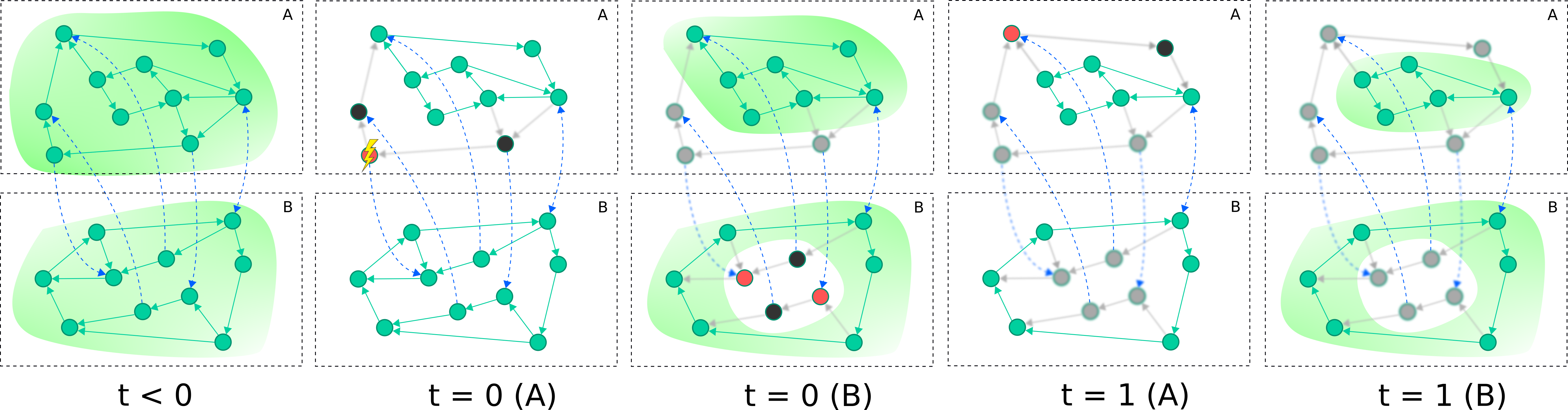}
        \put(0,0){\bf{(a)}}
    \end{overpic}} \\
    \subfloat{\begin{overpic}[width=0.6\linewidth]{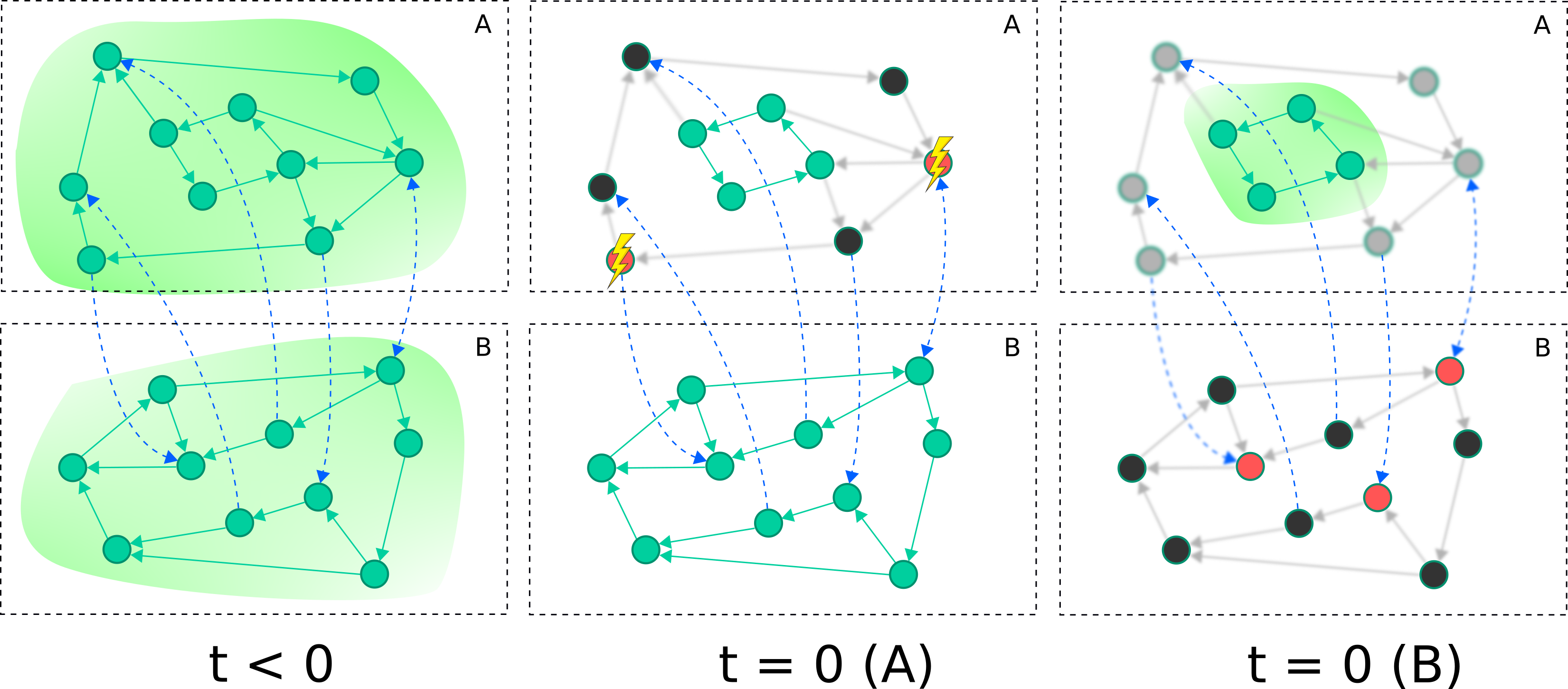} 
        \put(0,0){\bf{(b)}}
    \end{overpic}}
    \caption{Cascading failures in two networks of size $N = 10$ and $q_A = q_B = 0.3$. At each time step,
    events occur first in network A and then propagate to network B. (a) At $t < 0$, the GSCC of each network
    is highlighted with a light green shadow. At $t = 0$, a node is struck by lightning and fails 
    ($1 - p = 0.1$). As a result, the two black nodes turn into finite clusters and fail, as they get 
    disconnected from the GSCC\textsubscript{A}. Two of the three A-failures propagate to B through 
    dependencies (dashed lines). These B-failures are highlighted in red. Then, network B fragments into
    GSCC\textsubscript{B} and finite clusters (black nodes), the latter of which cause subsequent failures in
    A. At $t = 1$, network A fragments and the cascade stops since failures from finite clusters in A cannot
    propagate to B. The size of the MGSCC is 
    $P_\infty = \big( P^A_{\infty} + P^B_{\infty} \big) = (0.5 + 0.6)/2 = 0.55$. (b) Initially, two node fail
    in network A ($1 - p = 0.2$) producing fragmentation into GSCC\textsubscript{A} and finite clusters (black
    nodes). Failures spread to B and completely eliminate the GSCC\textsubscript{B}. The cascade stops at 
    $t = 0$ as finite clusters in B do not provide support to any functional A-node. The size of the MGSCC is 
    $P_\infty = 0$, since $P^B_{\infty} = 0$.} 
    \label{scheme}
\end{figure}

At time step $t = 0$, we start the cascading failures by assuming that a random fraction $1 - p$ of nodes in
layer A fail, and we remove these nodes. This causes the fragmentation into GSCC\textsubscript{A}, which 
remains functional, and finite clusters (FC\textsubscript{A}), which we consider to malfunction as they stop
receiving enough support from their own network because they lose connection with the GSCC\textsubscript{A}. 
Thus, finite clusters are removed too. A fraction $q_B$ of the total removed nodes in A cease to provide 
support to the same fraction of nodes in layer B, causing its fragmentation into GSCC\textsubscript{B} and 
FC\textsubscript{B}. Again, we assume that finite clusters in B fail. Hereafter, the failures that will pass
from one network to the other will only be due to finite clusters. In Fig.~\ref{scheme}, we show a scheme of
the process for a small system. At the end of the process, $P^A_{\infty}$ and $P^B_{\infty}$ are the 
fractions of nodes in GSCC\textsubscript{A} and GSCC\textsubscript{B}, respectively. The union of these 
components make up the MGSCC, of size $P_{\infty} = (P^A_{\infty} + P^B_{\infty})/2$, as long as both 
components have a finite size different from zero and they are interconnected through dependency links 
(Fig.~\ref{scheme} (a)). Otherwise, $P_{\infty} = 0$ (Fig.~\ref{scheme} (b)). If the cascade evolves without 
any kind of intervention, there is a critical point $p_c$ below which the interdependent directed networks 
collapse, i.e. $P_{\infty} = 0$ for $p \leq p_c$~\cite{liu-16}. As observed in~\cite{liu-16}, and similar to
non-directed interdependent networks~\cite{par-10}, the transition at $p_c$ changes from second-order to 
first-order as the coupling between networks is decreased.

Therefore, in order to mitigate the damage in the system, or either delay or avoid its collapse, we
immediately implement a recovery strategy in both networks. The strategy consists in repairing, with
probability $\gamma$, a subset of nodes from the ``contour'' or perimeter C\textsubscript{i} of 
GSCC\textsubscript{i}, $i = A, B$. A node from layer $i$ belongs to C\textsubscript{i} if it is a failure
and has, at least, one incoming and one outgoing link that lead to GSCC\textsubscript{i}. The contour
nodes that we consider for repairing may fulfill any of the following conditions:
\begin{enumerate}
    \item Nodes which have no dependencies and that do not support any node in the other network.
    \item Nodes with dependencies, for which we also check if their dependencies are contour nodes in the
    other layer. If that is the case, we repair both nodes. If the second node belongs to the GSCC of its
    corresponding layer, then we just recover the first node.
    \item Nodes which do not have dependencies but support nodes in the other network, excluding the case
    where the second node is a contour node. In this case, we only repair the first node.
\end{enumerate}
The restoring of a node implies that it turns back to a functional state and all of its links with the 
GSCC become available again. Note that, in this way, we can preserve the original degree distributions in
both networks, since we use already existing links to recover contour nodes. Therefore, we are able to 
map our model directly with the process of node percolation~\cite{dor-01,liu-16}.

Finally, after applying the strategy, we increase time in one unit and propagate the failures
from the finite clusters FC\textsubscript{B} back to network A, taking into account the 
fraction $q_A$ of nodes in A that have dependencies. The time step $t > 0$ includes the failure
of nodes in A due to dependencies with nodes in FC\textsubscript{B} at $t - 1$, the subsequent
propagation of failures from FC\textsubscript{A} to layer B, and the implementation of the 
recovery strategy in both layers. This process of cascading failures with recovery continues 
until a stationary state is reached, which depends on the values of $p$, $q_A$, $q_B$, and
$\gamma$, where there are no more failures and neither contour nodes available to repair.

\section{Analytical results and simulations}

In the following, we develop a theoretical approach to describe the model introduced in the previous section
by establishing an equivalent process of node percolation and using the generating functions
$G^{\ in}_{0i}(x) = \sum_{k_i^{\ in}} P^{\ in}_i(k_i^{\ in}) x^{k_i^{\ in}}$ and 
$G^{\ out}_{0i}(x) = \sum_{k_i^{\ out}} P^{\ out}_i(k_i^{\ out}) x^{k_i^{\ out}}$, $i = A, B$, which generate
the self and mutually uncorrelated degree distributions $P^{\ in}_i(k_i^{\ in})$ and 
$P^{\ out}_i(k_i^{\ out})$, respectively. Let $p_i(t)$ be the effective fraction of remaining nodes in network
$i$ at time $t$, before applying the recovery strategy. Given that the in-degree is not correlated with the 
out-degree, the fraction of nodes belonging to the GSCC\textsubscript{i} is~\cite{dor-01}
\begin{equation*}
    P_{\infty}^i(t) = p_i(t) \, g_i(p_i(t)), \quad g_i(x) = 
    \big( 1 - G^{\ in}_{0i}(1 - f_{\infty i}^{\ in}(x)) \big) 
    \big( 1 - G^{\ out}_{0i}(1 - f_{\infty i}^{\ out}(x)) \big),    
\end{equation*}
where $f_{\infty i}^{\ in}$ ($f_{\infty i}^{\ out}$) represents the probability of starting from a 
randomly chosen link in network $i$ and moving only against (along) the edge directions to arrive to the
S\textsubscript{i}\textsuperscript{out} (S\textsubscript{i}\textsuperscript{in}) component (remember 
Fig.~\ref{dir-net-comps}). The probabilities $f_{\infty i}^{\ in}$ and $f_{\infty i}^{\ out}$ satisfy the
transcendental equations 
\begin{eqnarray*}
    f_{\infty i}^{\ in} &=& x \big( 1 - G^{\ in}_{1i}(1 - f_{\infty i}^{\ in}) \big), \\
    f_{\infty i}^{\ out} &=& x \big( 1 - G^{\ out}_{1i}(1 - f_{\infty i}^{\ out}) \big),    
\end{eqnarray*}
where $G^{\ in}_{1i}$ and $G^{\ out}_{1i}$ are the generating functions of the number of 
incoming links arriving at a node reached by moving against the direction of a randomly chosen
link and the number of outgoing links leaving a node reached by moving along the direction of a
randomly chosen link, respectively~\cite{dor-01}.

Recall that the cascading failures start, at time $t = 0$, with the random removal of a fraction 
of A-nodes, denoted by $1 - p$. The remaining fraction of nodes in A is $p_A(0) = p$, where a
fraction $FC_A(0) = p_A(0) - P_{\infty}^A(0) = p (1 - g_A(p))$ of those nodes corresponds to 
finite clusters. Then, the remaining fraction of nodes in B is 
$p_B(0) = 1 - q_B \big( 1 - P_{\infty}^A(0) \big) = 1 - q_B \big( 1 - pg_A(p) \big)$, since
initial failures in A and finite clusters both affect network B, at the beginning, through
dependencies. The corresponding size of the finite clusters in B is 
$FC_B(0) = p_B(0) - P_{\infty}^B(0)$. At this point of each time step, once the cascade 
propagates from A to B, we apply the recovery strategy described in the {\it Model} section.

In order to implement our strategy, we need to find an expression for the fractions $C_A(t)$ and
$C_B(t)$ of nodes that belong to the contour of GSCC\textsubscript{A} and GSCC\textsubscript{B},
respectively. Take, for instance, the fraction $1 - p_A(t)$ that accounts for failed nodes in 
network A due to node percolation, at time $t$. We should multiply this by the probability 
$\big( 1 - G^{\ in}_{0A}(1 - P_{\infty}^A(t)) \big)$ that a randomly chosen node has, at least,
one incoming link from a node that belongs to the GSCC\textsubscript{A} and the probability 
$\big( 1 - G^{\ out}_{0A}(1 - P_{\infty}^A(t)) \big)$ that it has, at least, one outgoing link
leading to a node in the GSCC\textsubscript{A}. A similar reasoning holds for network B, 
yielding the fractions of contour nodes in each layer
\begin{eqnarray}
    \label{CA}
    C_A(t) &=& (1 - p_A(t)) \big( 1 - G^{\ in}_{0A}(1 - P_{\infty}^A(t)) \big)
    \big( 1 - G^{\ out}_{0A}(1 - P_{\infty}^A(t)) \big), \\
    \label{CB}
    C_B(t) &=& (1 - p_B(t)) \big( 1 - G^{\ in}_{0B}(1 - P_{\infty}^B(t)) \big)
    \big( 1 - G^{\ out}_{0B}(1 - P_{\infty}^B(t)) \big).
\end{eqnarray}
Then we update the relative sizes of GSCC\textsubscript{A} and GSCC\textsubscript{B} due to the
recovery of a subset of contour nodes from each network, with probability $\gamma$,
\begin{align}
    \label{PA_new}
    \nonumber \overline{P_{\infty}^A}(t) = P_{\infty}^A(t) \, + 
    \gamma &\Bigg[ (1 - q_A)(1 - q_B)C_A(t) + \frac{q_A (1 - q_B) C_A(t) C_B(t)}{1 - P_{\infty}^A(t)} +
    \frac{q_B C_A(t) C_B(t)}{1 - P_{\infty}^A(t)} + \\
    &+ \frac{(1 - q_A) q_B C_A(t) \big(1 - P_{\infty}^B(t) - FC_B(t) - C_B(t)\big)}
    {1 - P_{\infty}^A(t)} \Bigg], \\
    \label{PB_new}
    \nonumber \overline{P_{\infty}^B}(t) = P_{\infty}^B(t) \, + 
    \gamma &\Bigg[ (1 - q_A) (1 - q_B) C_B(t) + \frac{q_A (1 - q_B) C_A(t) C_B(t)}{1 - P_{\infty}^A(t)} +
    \frac{q_B C_A(t) C_B(t)}{1 - P_{\infty}^A(t)} + \\
    &+ \frac{(1 - q_B) q_A C_B(t) \big(1 - P_{\infty}^A(t) - C_A(t) \big)}
    {1 - P_{\infty}^A(t)} \Bigg],
\end{align}
where $\overline{P_{\infty}^i}(t)$ is the new relative size of GSCC\textsubscript{i}, $i = A, B$.
Before continuing with the next step, we look at Eq.~(\ref{PA_new}) for layer A to understand the
meaning of each term inside the square brackets, for which we include useful diagrams in 
Fig.~\ref{salv-nodes-scheme}: 
\begin{figure}
    \centering
    \subfloat{\begin{overpic}[width=\linewidth]{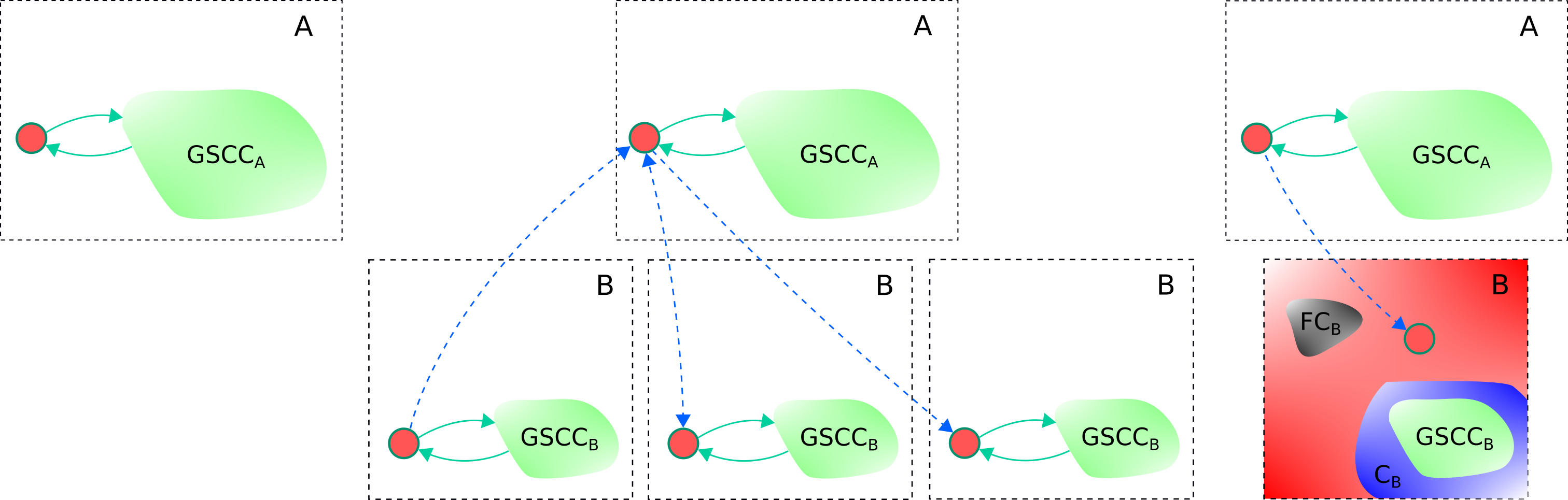}
        \put(1.5,29){\bf{(a)}}
        \put(40.5,29){\bf{(b)}}
        \put(79,29){\bf{(c)}}
    \end{overpic}}
    \caption{Scheme of the salvageable contour nodes. We focus on the nodes that can be recovered in
    network A, according to Eq.~(\ref{PA_new}). Note that single red nodes, both in A and B, belong 
    to the corresponding contour of each network, as they have one incoming and one outgoing link 
    that connects them to the GSCC (they may have more connections with the GSCC). In (a), we show
    the contour nodes in network A that do not have dependencies and do not support any B-nodes. The 
    diagram in (b) shows the different possibilities for contour nodes in A connected through
    dependencies with contour nodes in B. The last two instances are represented by the term 
    $q_B C_A(t) C_B(t)$ in Eq.~(\ref{PA_new}). In (c), we show the case in which contour nodes in A 
    provide support to failed B-nodes.}
    \label{salv-nodes-scheme}
\end{figure}
\begin{itemize}
    \item $(1 - q_A) (1 - q_B) C_A(t)$: Nodes belonging to the contour of GSCC\textsubscript{A} that
    have no dependencies on B-nodes and do not support any B-nodes (Fig.~\ref{salv-nodes-scheme} (a)).
    \item Nodes in C\textsubscript{A} that are interconnected with nodes in C\textsubscript{B} 
    (Fig.~\ref{salv-nodes-scheme} (b)):
    \begin{itemize}
        \item $q_A (1 - q_B) C_A(t) C_B(t)$: Nodes in C\textsubscript{A} that depend on nodes from
        C\textsubscript{B} but do not provide support to them. When the strategy is not applied, this
        kind of dependencies do not exist because contour nodes in B always depend on failures from
        A that may belong to C\textsubscript{A} or not. However, when the strategy is applied, the
        recovery of contour nodes may turn some failures in the perimeter of C\textsubscript{A} and
        C\textsubscript{B} into new contour nodes, which are considered for repair in the next time
        step. Thus, while this case is absent at $t = 0$, we must consider it for any other time.
        \item $q_B C_A(t) C_B(t)$: Nodes in C\textsubscript{A} that provide support to nodes in 
        C\textsubscript{B}, whether or not the C\textsubscript{A} nodes depend on the same 
        C\textsubscript{B} nodes.
    \end{itemize}
    Note that these two terms involve contour nodes from both networks. As nodes are recovered both in 
    A and B, the terms are present as well in Eq.~(\ref{PB_new}) for network B.
    \item $(1 - q_A)q_B C_A(t) \big(1 - P_{\infty}^B(t) - FC_B(t) - C_B(t) \big)$: Nodes from 
    C\textsubscript{A} that do not depend on B-nodes but provide support to failures from network B 
    (Fig.~\ref{salv-nodes-scheme} (c)). Note that we subtract the finite clusters $FC_B(t)$ from the
    total failures $1 - P_{\infty}^B(t)$ because, by definition, finite clusters in B are functional 
    right after failures propagate from A to B, and then fail due to being disconnected from the 
    GSCC\textsubscript{B}. Thus, they cannot depend on failed A-nodes. We also exclude nodes from
    C\textsubscript{B} as we have already considered this case in the term $q_B C_A(t) C_B(t)$.  
\end{itemize}
To complete the description, note that all terms inside the square brackets involving dependencies
are conditioned by the probability that dependencies can be established between the corresponding nodes,
which is in all cases $1 - P_{\infty}^A(t)$. The possibilities for network B, reflected on 
Eq.~(\ref{PB_new}), are similar to those in A with an exception in the last term,
$(1 - q_B)q_A C_B(t) \big(1 - P_{\infty}^A(t) - C_B(t) \big)$, where it is not necessary to subtract the
finite clusters $FC_A(t)$ to the total failures $1 - P_{\infty}^A(t)$ because finite clusters in A are
conformed before $C_B(t)$ and, thus, nodes in FC\textsubscript{A} can depend on failed nodes.

Once we update the sizes of GSCC\textsubscript{A} and GSCC\textsubscript{B}, we need to compute the
effective fractions of remaining nodes in each network after the application of the strategy, 
$\overline{p_A}(t)$ and $\overline{p_B}(t)$, by solving the transcendental equations
\begin{eqnarray*}
    \overline{p_A}(t) g_A(\overline{p_A}(t)) = \overline{P_{\infty}^A}(t), \\
    \overline{p_B}(t) g_B(\overline{p_B}(t)) = \overline{P_{\infty}^B}(t).    
\end{eqnarray*}
After this step, we propagate the cascade back to network A through failures in FC\textsubscript{B} 
and we start a new time step. The remaining fractions of nodes before the implementation of the recovery 
strategy are, at time $t > 0$,
\begin{eqnarray}
    \label{pA_eff} 
    p_A(t) &=& \overline{p_A}(t - 1) \left( 1 - \frac{FC_B(t - 1) \left[ q_A q_B + 
    q_A(1 - q_B) \overline{P_{\infty}^A}(t - 1) \right]}
    {1 - q_B \big(1 - \overline{P_{\infty}^A}(t - 1) \big)} \right), \\
    \label{pB_eff}
    p_B(t) &=& \overline{p_B}(t - 1) \left( 1 - \frac{FC_A(t) \left[ q_A q_B + 
    q_B(1 - q_A) \overline{P_{\infty}^B}(t - 1) \right]}
    {1 - q_A \big(1 - \overline{P_{\infty}^B}(t - 1)\big)} \right),
\end{eqnarray}
where 
\begin{eqnarray}
    \label{FCA} 
    FC_A(t) &=& \overline{P_{\infty}^A}(t - 1) - FC_B(t - 1) \left[ q_A q_B + 
    q_A(1 - q_B) \overline{P_{\infty}^A}(t - 1) \right] - P_{\infty}^A(t), \\
    \label{FCB}
    FC_B(t) &=& \overline{P_{\infty}^B}(t - 1) - FC_A(t) \left[ q_A q_B + 
    q_B(1 - q_A) \overline{P_{\infty}^B}(t - 1) \right] - P_{\infty}^B(t),
\end{eqnarray}
are the relative sizes of the finite clusters in layer A and B, respectively. In 
Appendix~\ref{app:fin-clu}, we make a brief comment on how we compute $FC_A(t)$ and $FC_B(t)$. Note that
the fractions of failures that we subtract from each GSCC in Eqs.~(\ref{FCA})-(\ref{FCB}) are the same
fractions that we use in Eqs.~(\ref{pA_eff})-(\ref{pB_eff}) to propagate failures between networks. 
However, to extend the failures outside the GSCC\textsubscript{A} and into $\overline{p_A}(t - 1)$, when
transmitting failures from FC\textsubscript{B} to network A in Eq.~(\ref{pA_eff}), we must normalize the
failures by leaving aside failed A-nodes that provide support to nodes in B, as these nodes could never be
connected to the FC\textsubscript{B}. This condition is represented by the denominator 
$1 - q_B \big(1 - \overline{P_{\infty}^A}(t - 1) \big)$. In a similar way, we obtain Eq.~(\ref{pB_eff}) 
for $p_B$.

\begin{figure}
    \centering
    \subfloat{\begin{overpic}[width=0.5\linewidth]{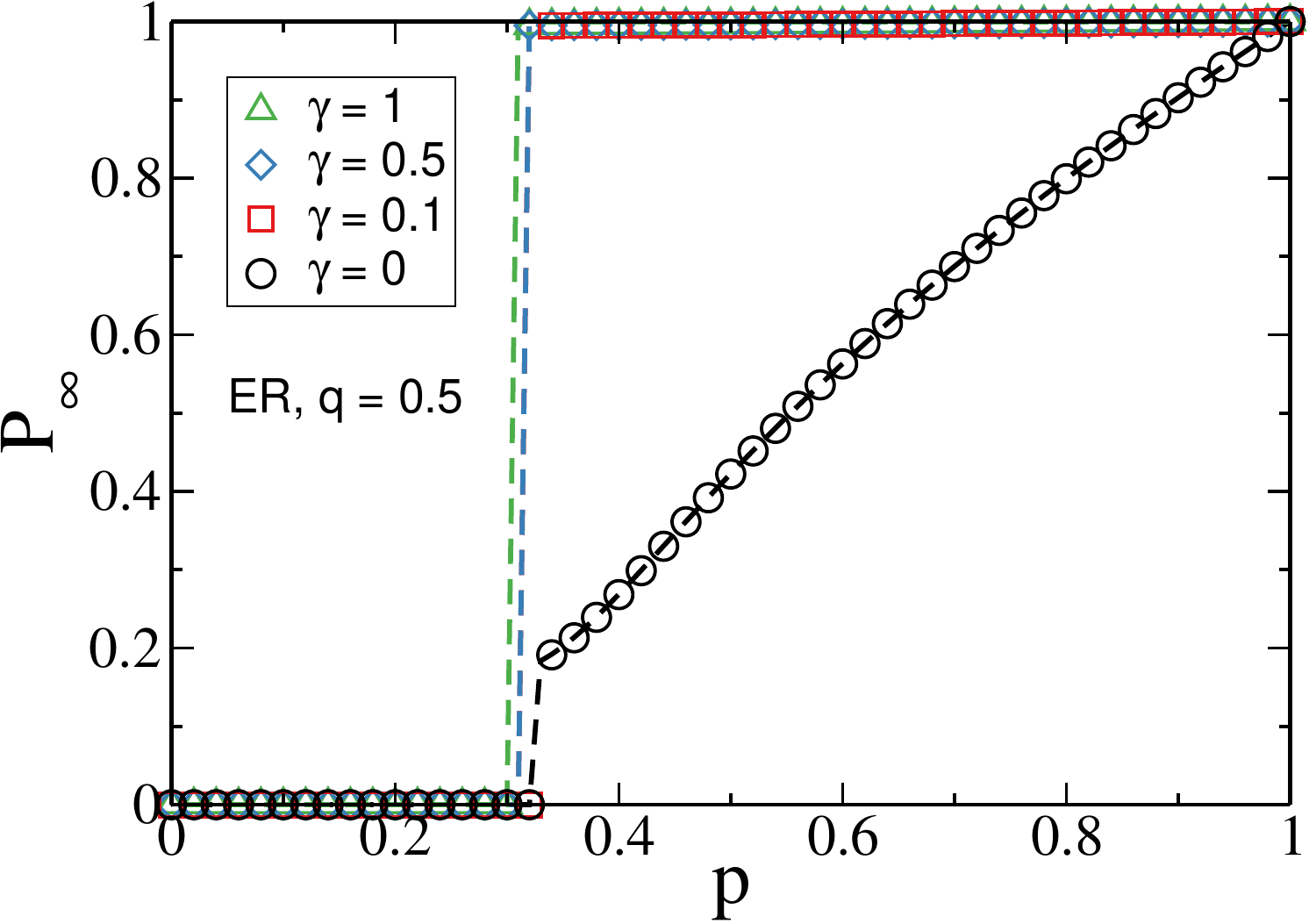}
        \put(0,0){\bf{(a)}}
    \end{overpic}}
    \subfloat{\begin{overpic}[width=0.5\linewidth]{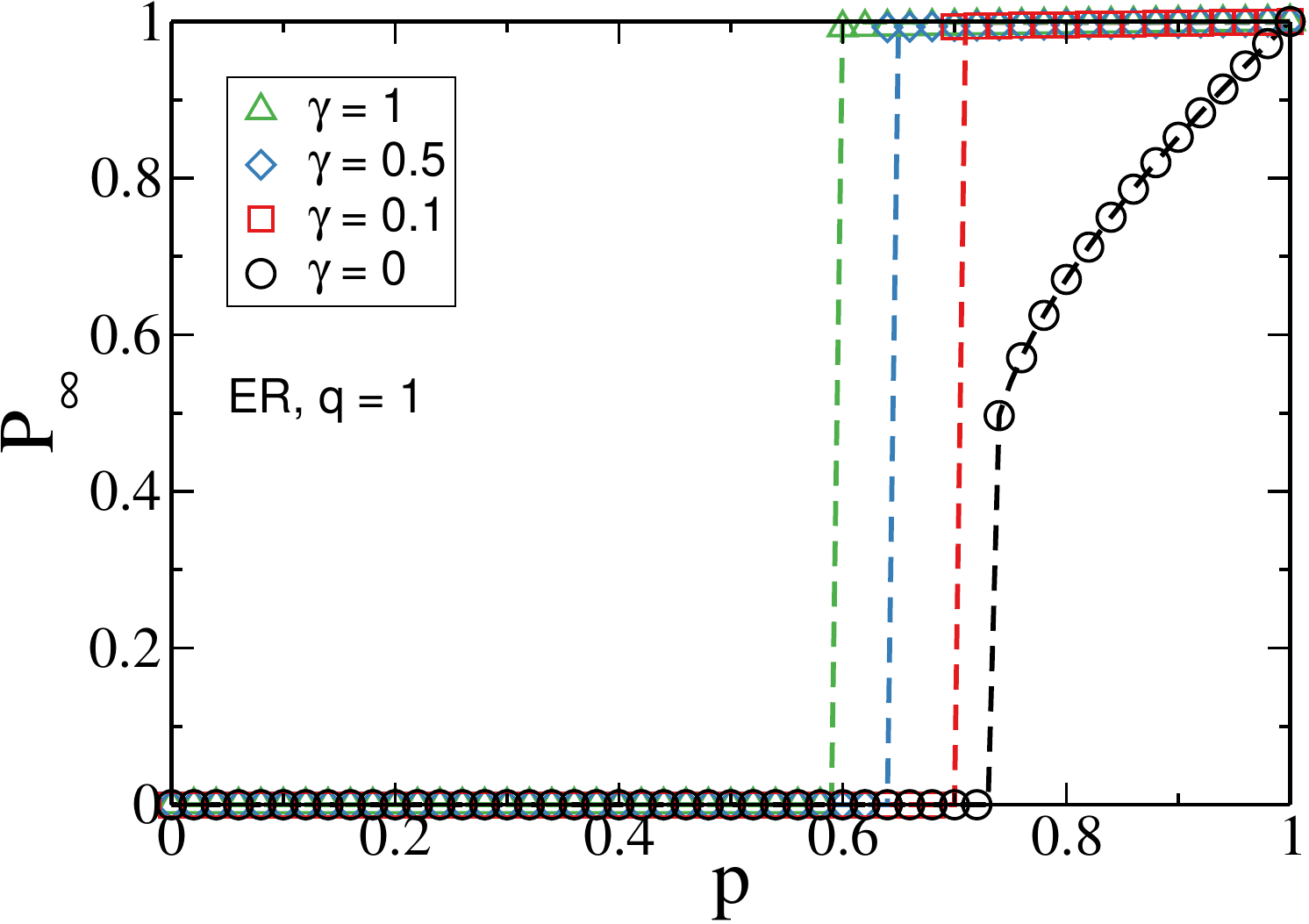} 
        \put(0,0){\bf{(b)}}
    \end{overpic}} \\
    \subfloat{\begin{overpic}[width=0.5\linewidth]{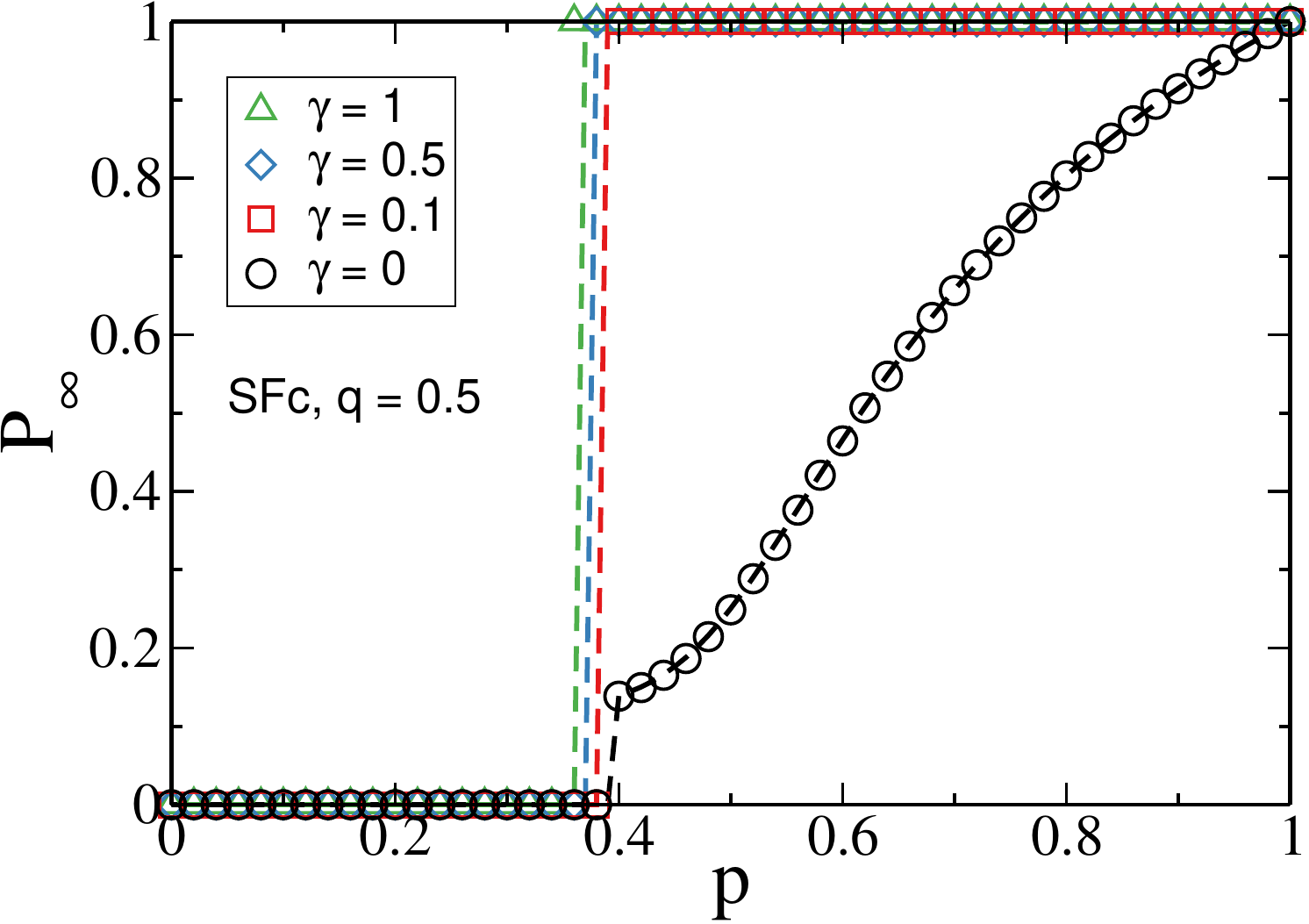} 
        \put(0,0){\bf{(c)}}
    \end{overpic}}
    \subfloat{\begin{overpic}[width=0.5\linewidth]{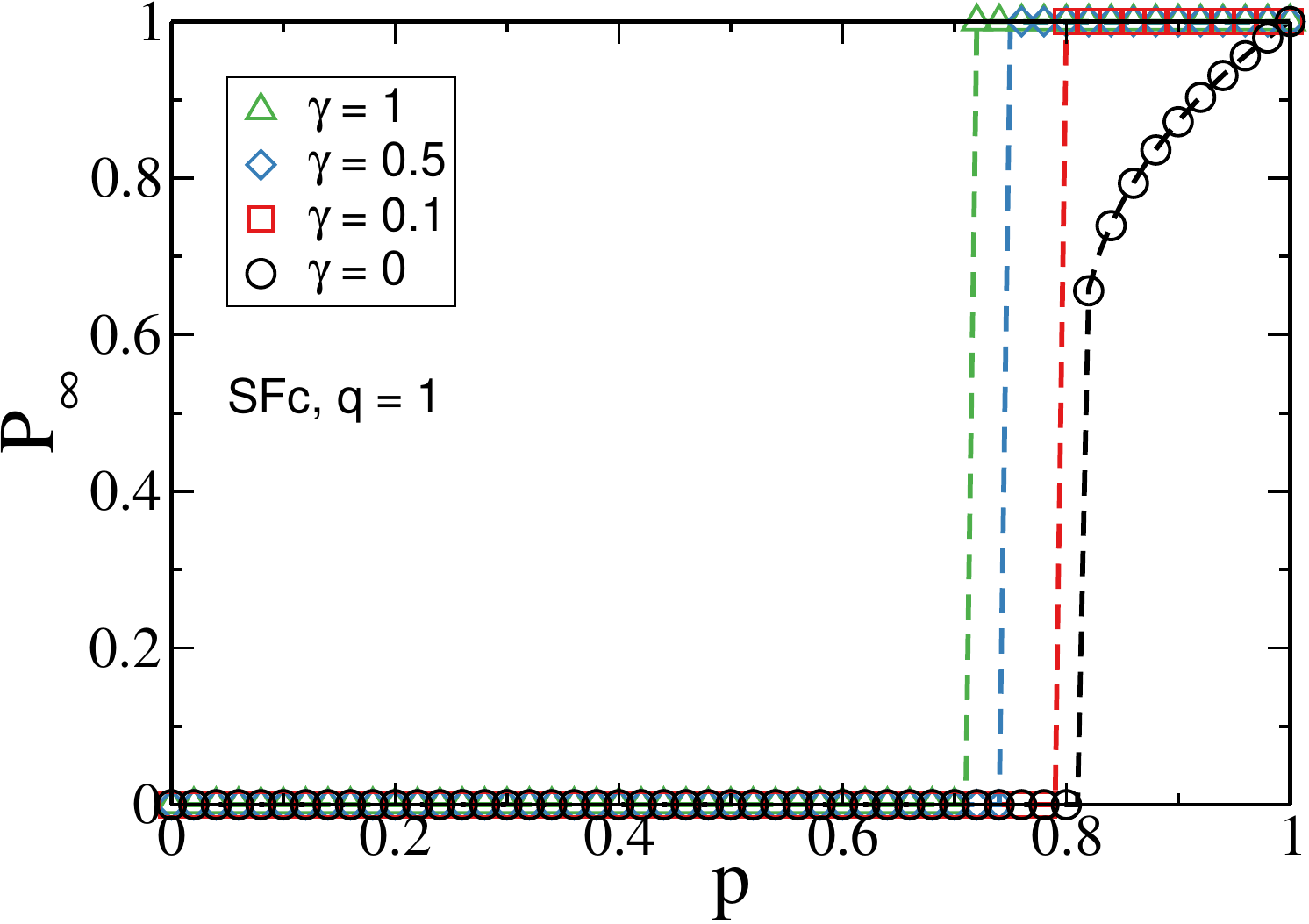} 
        \put(0,0){\bf{(d)}}
    \end{overpic}}
    \caption{Outcome of the recovery strategy. The analytical results (dashed lines) of
    the fraction of nodes in the MGSCC, $P_{\infty}$, as function of the fraction of 
    initial remaining nodes $p = p_A(0)$ in network A show a good agreement with the simulations
    (symbols) for the different values of $\gamma$ and $q$. We include the case $\gamma = 0$
    as a reference. In (a) and (b), $P(k)$ corresponds to an ER network with $\valmed{k} = 4$,
    $k_{min} = 1$, and $k_{max} = 20$. In (c) and (d), we use a SF distribution with 
    $\lambda = 2.35$, exponential cutoff $k_{cut} = 50$, $k_{min} = 2$, and $k_{max} = \sqrt{N}$,
    which yields $\valmed{k} \approx 4$. In all cases, for the simulations we build
    networks of size $N = 10^6$ and we average results over $10^3$ realizations.} 
    \label{Pinf}
\end{figure}

\begin{figure}
    \centering
    \subfloat{\begin{overpic}[width=0.5\linewidth]{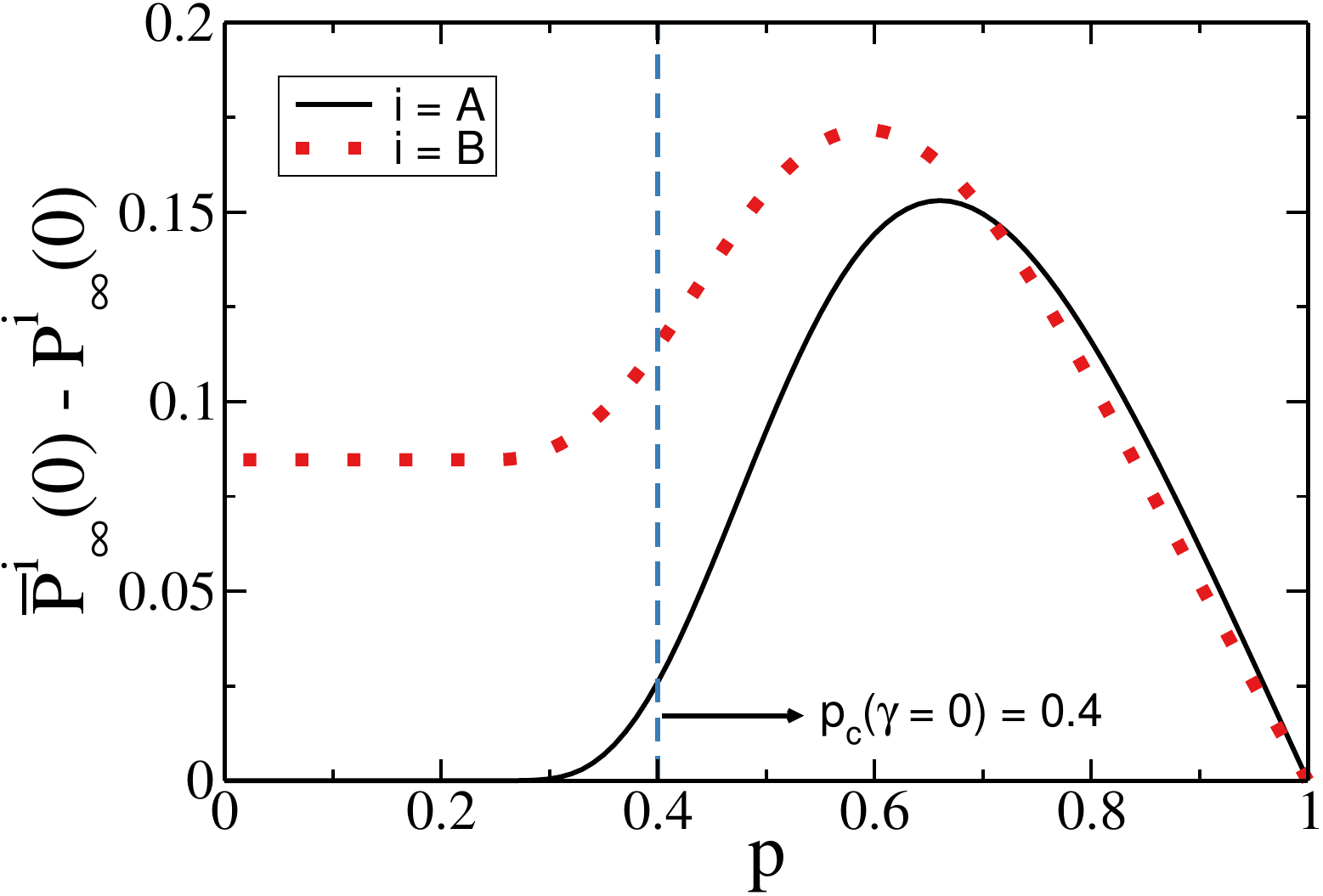}
        \put(0,0){\bf{(a)}}
    \end{overpic}}
    \subfloat{\begin{overpic}[width=0.5\linewidth]{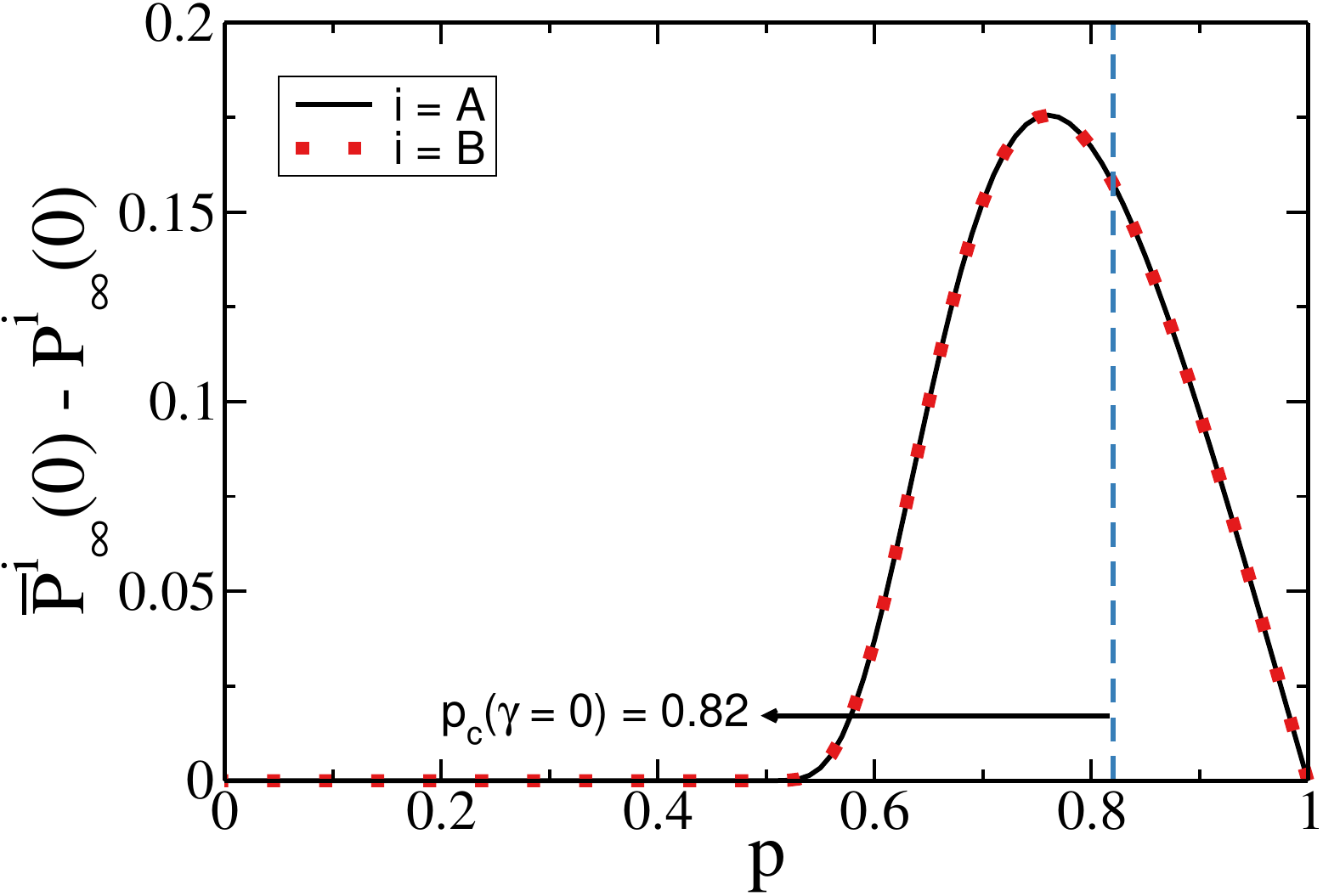} 
        \put(0,0){\bf{(b)}}
    \end{overpic}}
    \caption{Fractions of salvageable contour nodes in each network, 
    $\overline{P_{\infty}^i}(0) - P_{\infty}^i(0)$, as functions of $p = p_A(0)$ for (a) 
    $q = 0.5$ and (b) $q = 1$. We compute these fractions from Eqs.~(\ref{PA_new})-(\ref{PB_new})
    with $\gamma = 1$, for the interdependent directed networks presented in Fig.~\ref{Pinf} (c)
    and (d), which present a SFc degree distribution $P(k)$. The dashed lines mark the
    corresponding critical points in the absence of the recovery strategy ($\gamma = 0$).} 
    \label{salv-nodes}
\end{figure}

Based on this analytical approach, we focus on studying the relative size of the MGSCC at the final stage
of the process, $P_{\infty}$, which is the principal magnitude to reflect the degree of robustness of the
interdependent directed networks. The MGSCC includes the A-nodes that belong to GSCC\textsubscript{A} and
the B-nodes in GSCC\textsubscript{B} when the two components are interconnected through dependencies that 
are set in both ways, i.e. there are nodes in GSCC\textsubscript{A} that depend on nodes from 
GSCC\textsubscript{B} and vice-versa. In this case, the relative size of the MGSCC is 
$P_{\infty} = (P_{\infty}^A + P_{\infty}^B)/2$, where $P_{\infty}^A$ and $P_{\infty}^B$ are the non-zero
relative sizes of GSCC\textsubscript{A} and GSCC\textsubscript{B} at the steady state, respectively.
Otherwise, each layer would be isolated from the other one and the system would lose its original 
integration and functionality, yielding $P_\infty = 0$. In Fig.~\ref{Pinf}, we show $P_{\infty}$ as a
function of $p$ for two interdependent directed networks with degree distributions 
$P^{\ in}_A(k_A^{\ in} = k) = P^{\ out}_A(k_A^{\ out} = k) = P^{\ in}_B(k_B^{\ in} = k) = 
P^{\ out}_B(k_B^{\ out} = k) \equiv P(k)$, with $P(k)$ corresponding to (a) an Erd\"{o}s-R\'{e}nyi (ER)
network and (b) a Scale Free network with exponential cutoff (SFc), and for different fractions of nodes 
with dependencies $q_A = q_B \equiv q$. We observe that the analytical results (dashed lines) are in well
agreement with the results from the simulations, which we represent with different symbols. 
Our results for $\gamma = 0$ are consistent with those in~\cite{liu-16} where, in the absence
of a recovery strategy, the critical point $p_c$ becomes smaller as the fraction of 
interdependent nodes $q$ decreases, which makes the system more robust to cascading failures
although less interconnected. This behavior is generalized, in our model, to the scenario where 
a recovery strategy is implemented ($\gamma > 0$). One of the most significant result that we can
extract from Fig.~\ref{Pinf} is that the robustness of the system improves when we implement the
recovery strategy and for increasing values of $\gamma$, to a greater or lesser extent. We 
observe that this effect on $p_c$ appears to be stronger in absolute terms when the fraction of
nodes with dependencies in both networks is $q = 1$ (Fig.~\ref{Pinf} (b) and (d)) although, in 
this case, the failures produce a greater damage to the system compared to the case where 
$q = 0.5$ (Fig.~\ref{Pinf} (a) and (c)). In order to understand this behavior, in 
Fig.~\ref{salv-nodes} we plot the initial fractions of salvageable contour nodes in each network, 
$\overline{P_{\infty}^i}(0) - P_{\infty}^i(0)$, $i = A, B$.
We observe that these fractions are larger for the fully interdependent directed networks, i.e.
for $q = 1$, at the critical point $p_c(\gamma = 0)$ (dashed line in Fig.~\ref{salv-nodes} (b)),
than for the case $q = 0.5$. Besides, for $q = 0.5$, salvageable contour nodes in A rapidly 
vanish as $p$ falls below $p_c(\gamma = 0)$ (dashed line in Fig.~\ref{salv-nodes} (a)), 
which limits the possibility of starting a recovery process that leads to the complete 
restoring of the system. The initial size of GSCC\textsubscript{i}, after the first step of
the cascade, plays a role too. In this case, it also yields bigger values for the fully
interdependent directed networks, but there could be intermediate scenarios where the outcome
of the balance between these magnitudes shows an opposite behavior, i.e. the decrease of the
interval where the strategy achieves to avoid the collapse of the system as $q$ becomes larger.

Moreover, we observe that the system either collapses ($P_{\infty} = 0$) or its functionality
is completely restored ($P_{\infty} = 1$) when we apply the recovery strategy, i.e. for
$\gamma > 0$. As the process evolves, if there is a step at which failures no longer propagate
and a MGSCC still survives, we continue with the recovery of the salvageable contour nodes in
each network until we are unable to repair any more. It is crucial for accomplishing the full
restoring of the system to repair contour nodes which do not have dependencies and do not
support other nodes, in addition to contour nodes which have dependencies and/or are supporting
nodes themselves. In this way, the critical point $p_c$ for $\gamma > 0$ corresponds to the
threshold for the abrupt transition between a collapsing phase, where $P_{\infty} = 0$ for 
$p \leq p_c$, and a completely functional phase, where $P_{\infty} = 1$ for $p > p_c$. This
characteristic is shared with the non-directed model~\cite{dim-15}. 

Finally, regarding the topology of the interdependent directed networks, Fig.~\ref{Pinf} shows
that the homogeneous systems (Fig.~\ref{Pinf} (a) and (b)), are more robust when compared to
heterogeneous systems (Fig.~\ref{Pinf} (c) and (d)) with the same average degree. The lower 
values for $p_c$ in systems with ER degree distributions are observed for all the values of
$q$ and also for all the values of the recovery probability $\gamma$. These results agree with
those obtained in non-directed models~\cite{bul-10,par-10,par-11-bis}.

\begin{figure}
    \centering
    \subfloat{\begin{overpic}[width=0.5\linewidth]{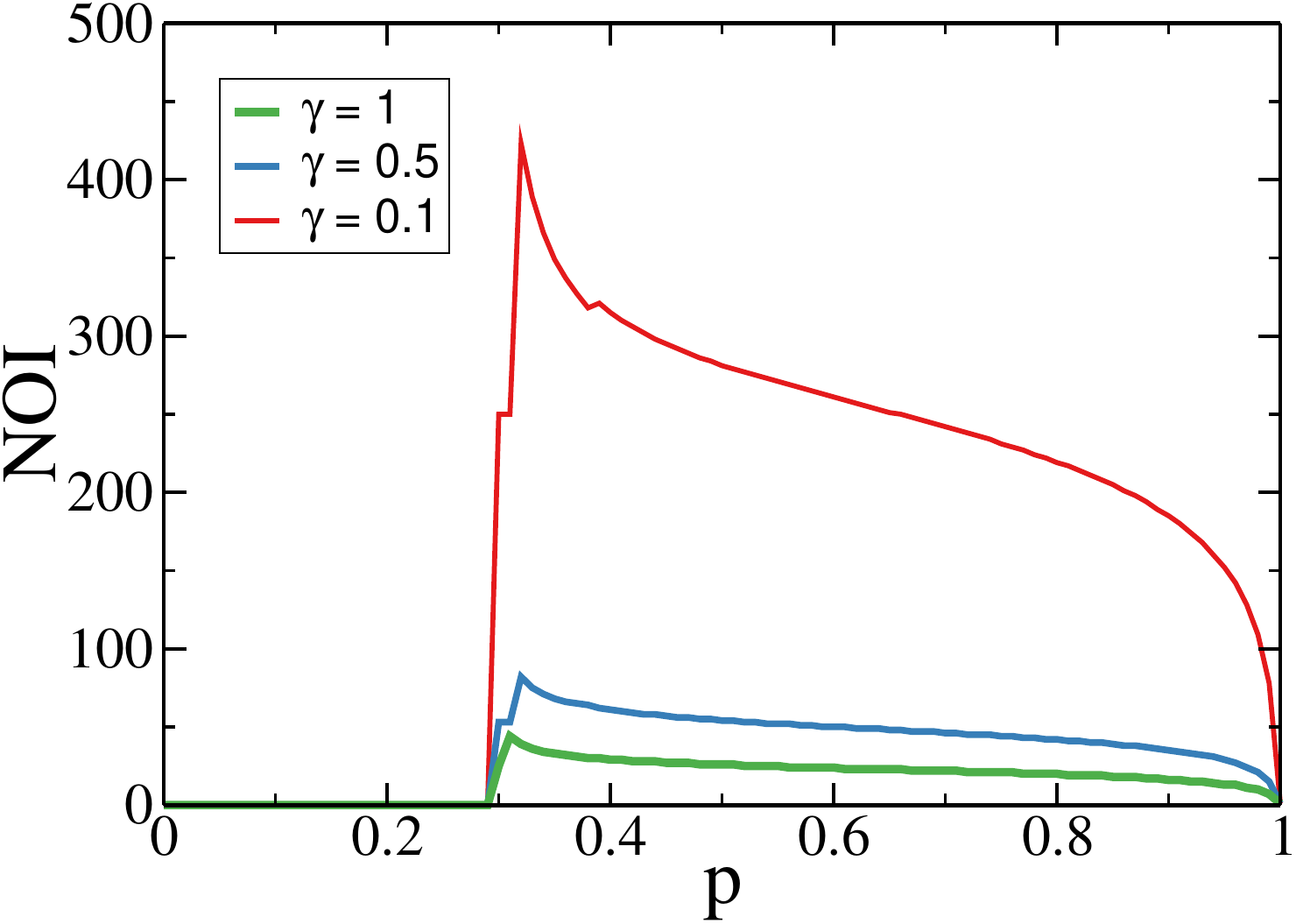}
        \put(0,0){\bf{(a)}}
    \end{overpic}}
    \subfloat{\begin{overpic}[width=0.5\linewidth]{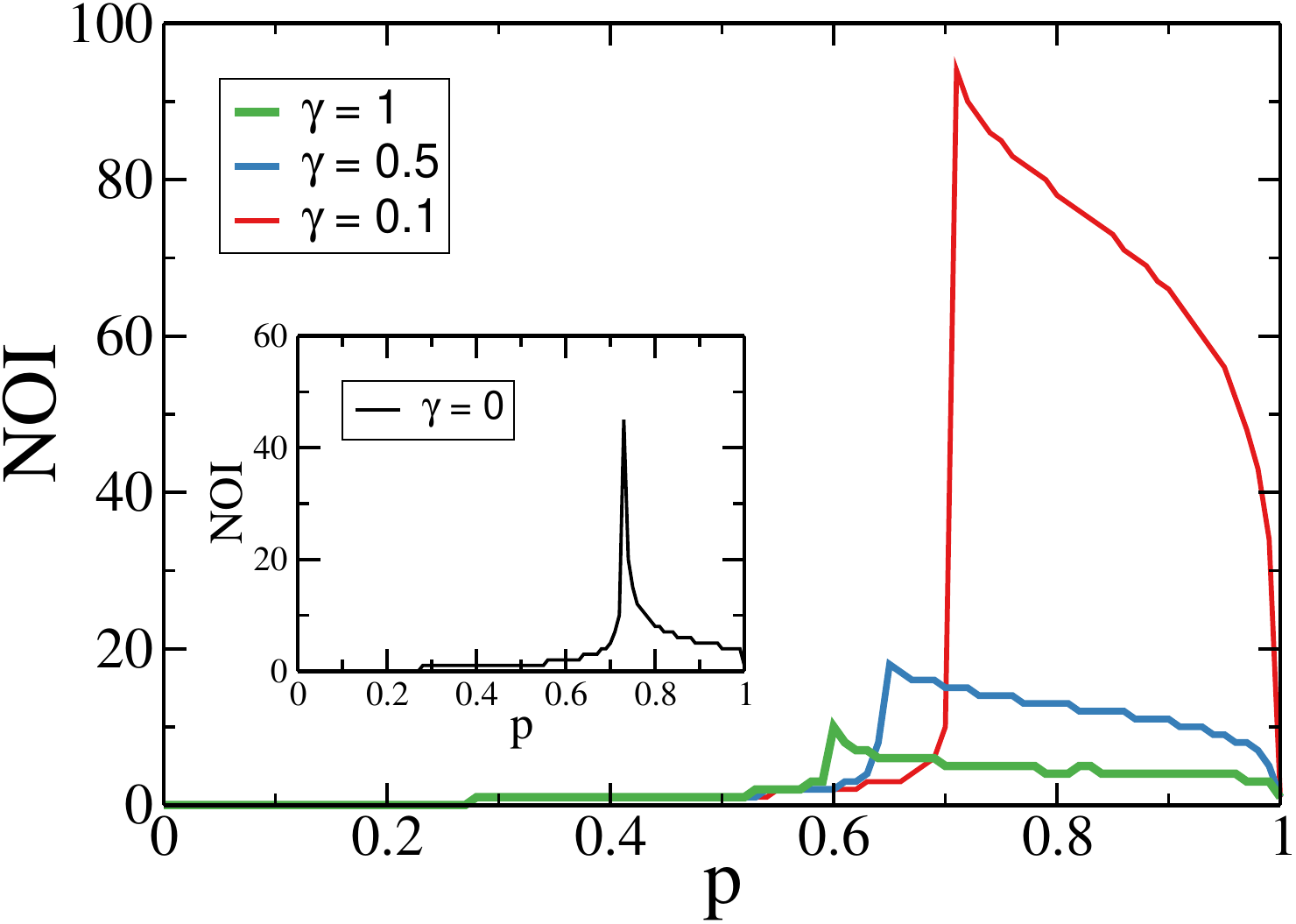} 
        \put(0,0){\bf{(b)}}
    \end{overpic}}
    \caption{Number of iterations (NOI) as function of $p = p_A(0)$, for the interdependent directed
    networks analyzed in Fig.~\ref{Pinf} (a) and (b), which present an ER degree distribution
    $P(k)$. Results in (a) correspond to $q = 0.5$, while in (b) $q = 1$. The thickness of the
    curves increase with the value of the recovery probability $\gamma$. We included an inset 
    in (b) for the case $\gamma = 0$, which shows the typical sharp peak at criticality.} 
    \label{NOI}
\end{figure}

In order to analyze the overall robustness of the system and the degree of difficulty with 
which it can be restored, we compute the phase diagrams in the plane $(p, \gamma)$ for the
interdependent directed networks analyzed above. A possible approach to do this is by means of
the number of iterations (NOI) that the system requires to reach the steady state. In other
words, the NOI accumulates the iterative steps (discrete time $t$), each one of which can 
comprehend the propagation of failures and the implementation of the recovery strategy or only 
one of these processes, until the system collapses or gets completely restored. In this way,
the NOI is a measure of the velocity of the process and it is known that, if no recovery
strategy is applied, it shows a peak in the vicinity of criticality~\cite{par-11-bis}, 
revealing a slow dynamic behavior. Thus, the number of steps necessary for the system to reach
the steady state increases significantly when $p$ is close to $p_c$, but away from criticality 
only a few steps are required. Therefore, the NOI can be used as an accurate measure for 
computing the critical threshold $p_c$ for each value of $\gamma$. In Fig.~\ref{NOI}, we show 
the NOI obtained from the theoretical approach for the homogeneous networks presented earlier 
(see Fig.~\ref{Pinf} (a) and (b)), which is enough to demonstrate the qualitative behavior of
this magnitude for our process.
In Figs.~\ref{NOI} (a) and (b) for $q = 0.5$ and $q = 1$, respectively, we see that there is a peak 
that signals the criticality if we implement the recovery strategy, although it is not as sharp as
for the case $\gamma = 0$ (see the inset in Fig.~\ref{NOI} (b)). Above the critical point, for 
$p > p_c$, the NOI appears to decrease in a constant manner, opposed to the abrupt way in which the
NOI increases as we get close to $p_c$ from below. In addition, the curves obtained for the different
values of $\gamma$ shows us that the number of steps that the system requires to reach the steady 
state decreases as the fraction of recovered nodes becomes larger. Measuring the position of the
maximum value in the NOI can involve extensive data analysis, as observed in the number of steps that
the process can last from Fig.~\ref{NOI} (a). For this reason, we appeal to an equivalent method for
analytically computing the phase diagrams, which is also used for the non-directed 
model~\cite{dim-15}. Given a fixed value of $p$, we run the theoretical model for decreasing values 
of the probability of recovery $\gamma$ until the size of the MGSCC falls to zero. We record the
value of $\gamma$ for which this occurs as $\gamma_c(p)$ and we find, in all cases, that this value
is the same that we obtain by measuring the peak of the NOI.

\begin{figure}
    \centering
    \subfloat{\begin{overpic}[width=0.5\linewidth]{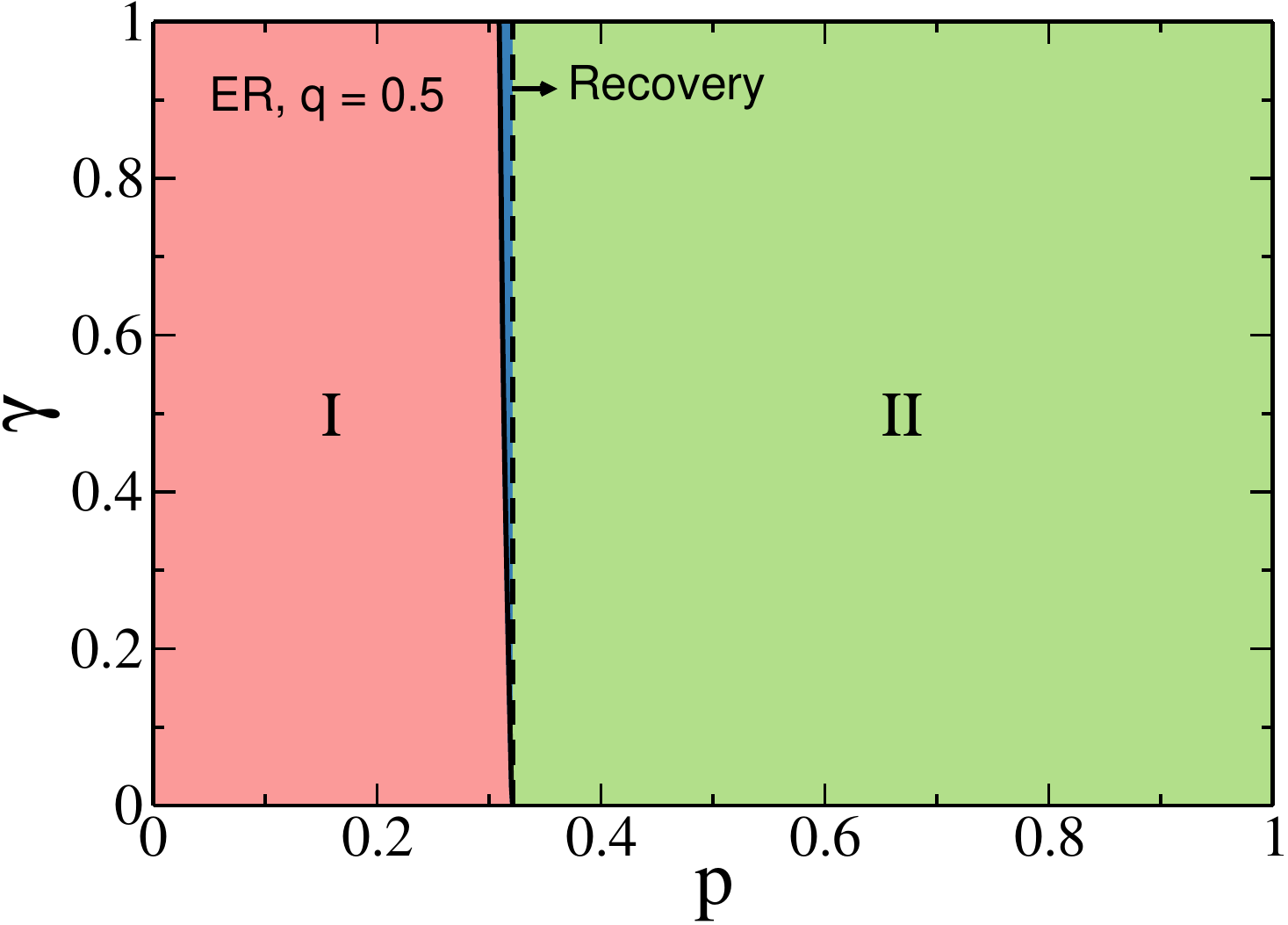}
        \put(0,0){\bf{(a)}}
    \end{overpic}}
    \subfloat{\begin{overpic}[width=0.5\linewidth]{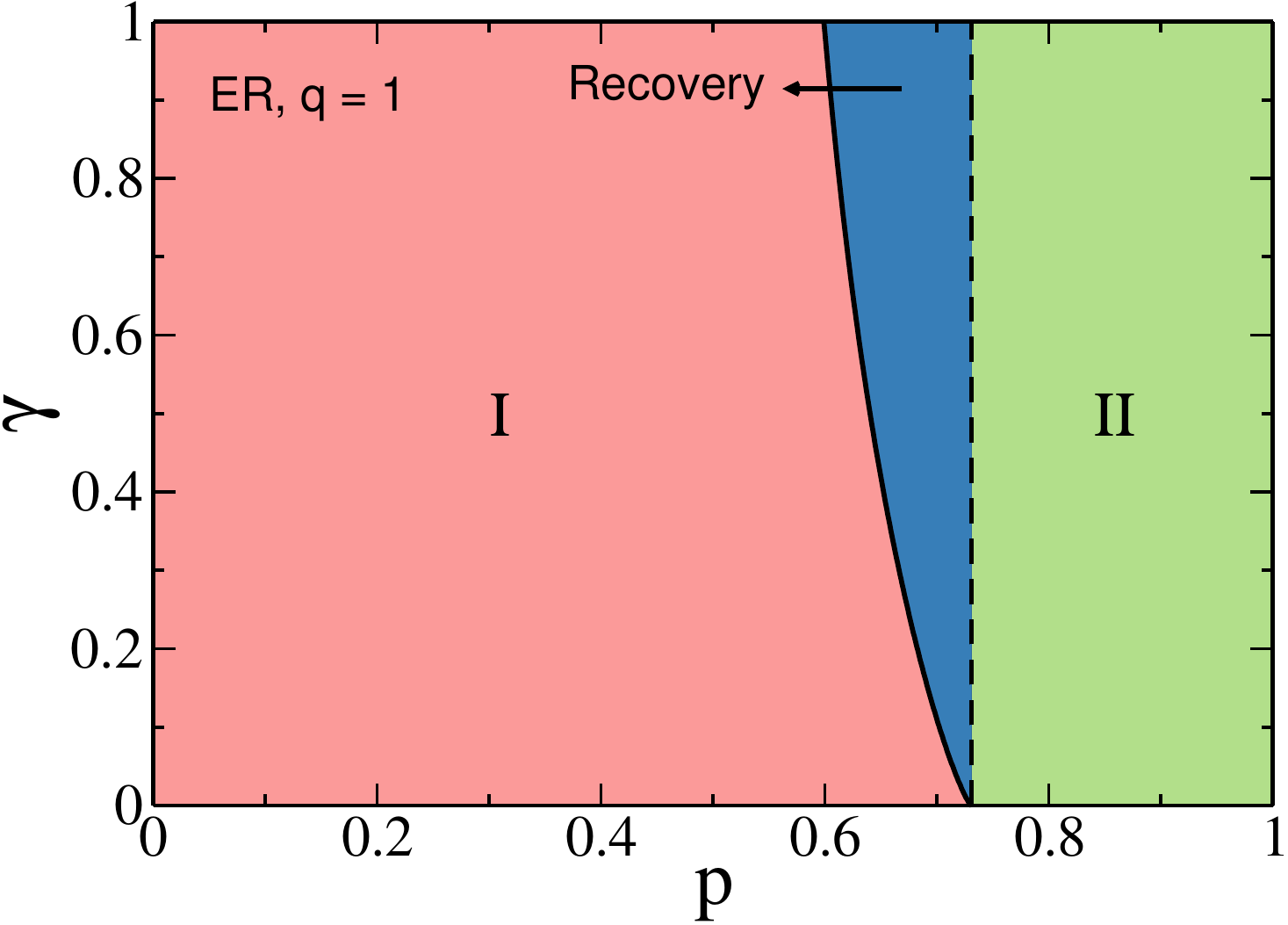} 
        \put(0,0){\bf{(b)}}
    \end{overpic}} \\
    \subfloat{\begin{overpic}[width=0.5\linewidth]{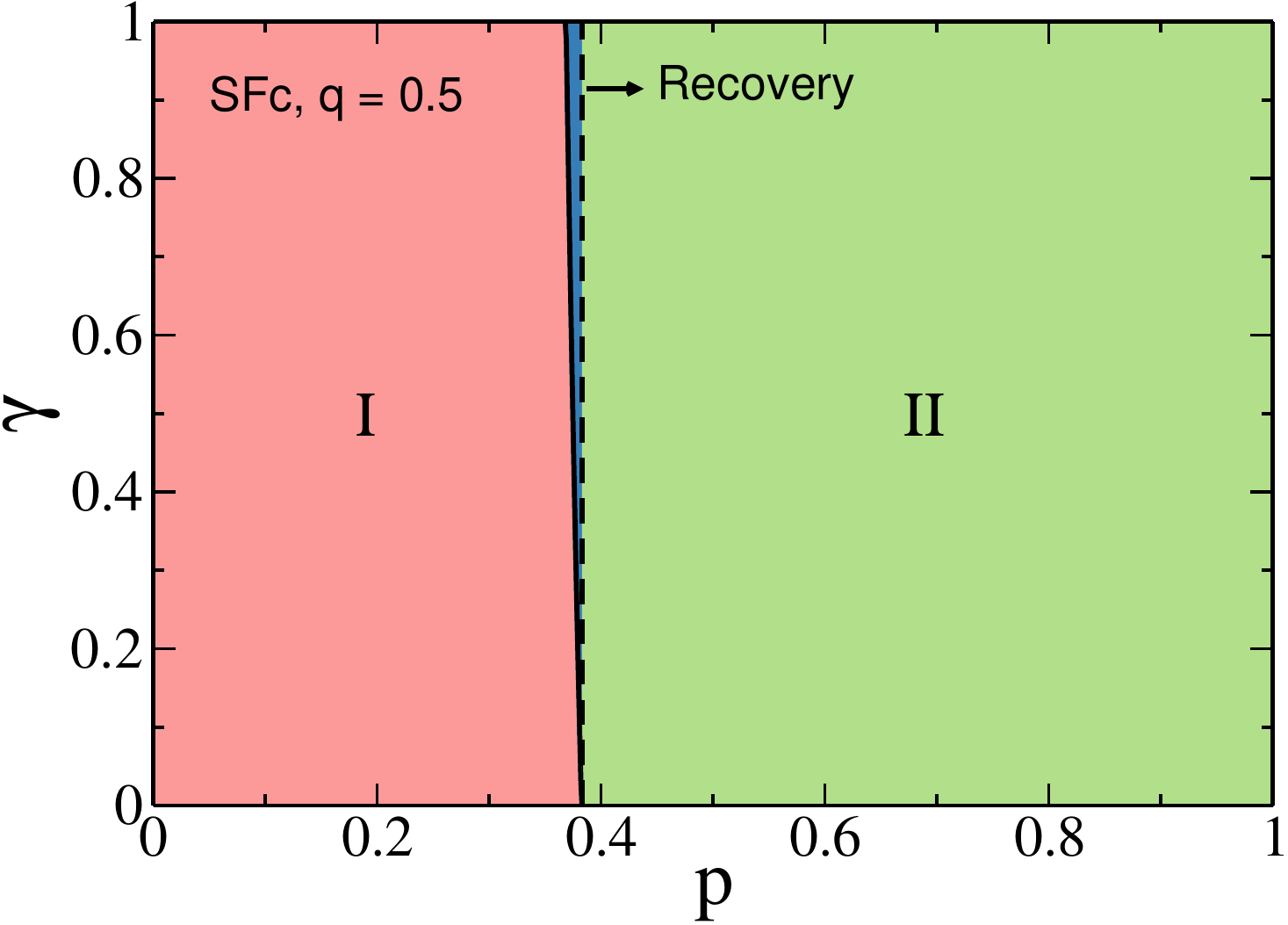} 
        \put(0,0){\bf{(c)}}
    \end{overpic}}
    \subfloat{\begin{overpic}[width=0.5\linewidth]{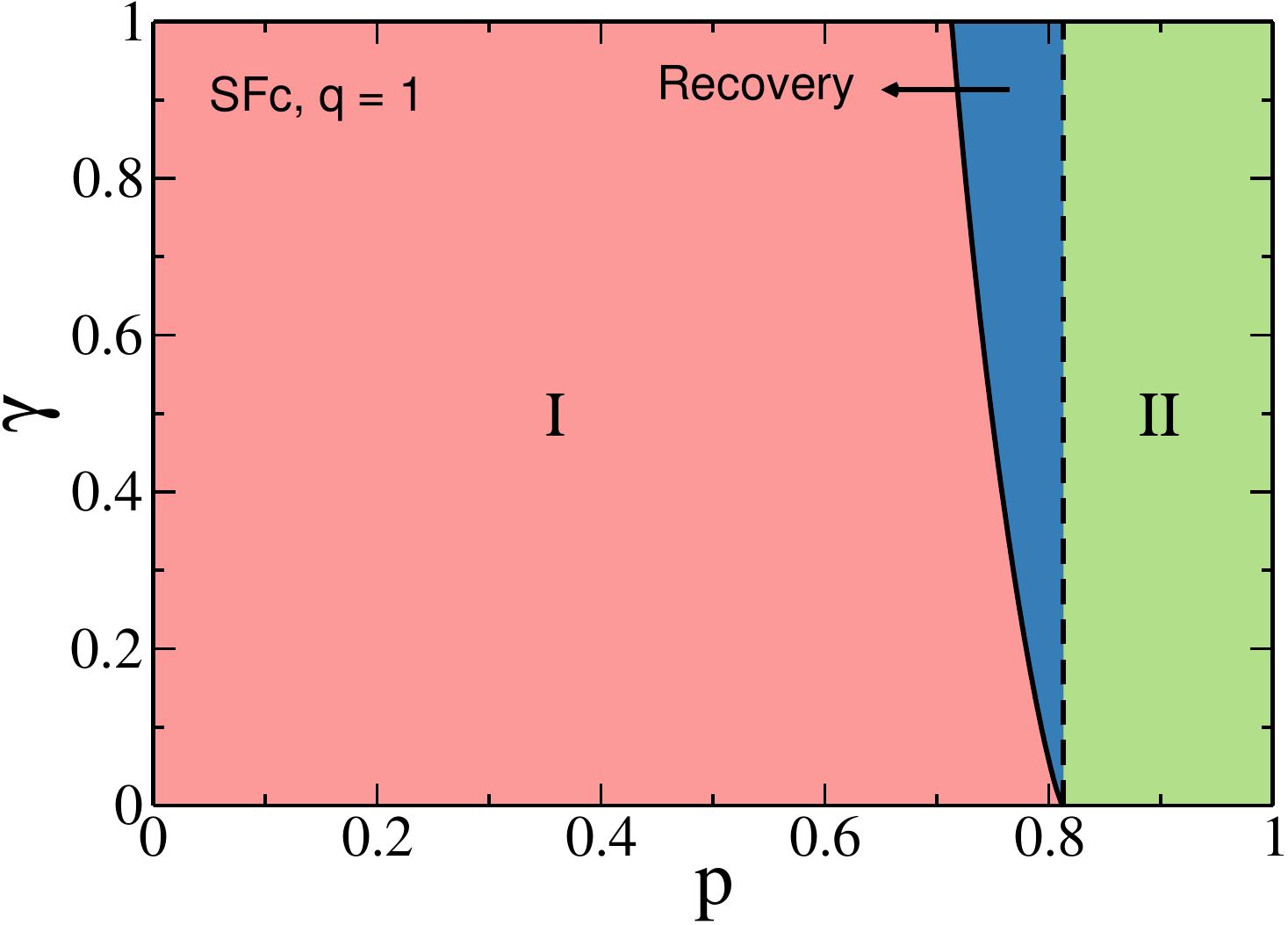} 
        \put(0,0){\bf{(d)}}
    \end{overpic}}
    \caption{Phase diagrams on the $(p, \gamma)$ plane, for the interdependent directed networks
    analyzed in Fig.~\ref{Pinf}. The solid lines represent the curves $\gamma_c(p)$ below which the
    system completely collapses (region I). The dashed lines correspond to the critical point in the
    absence of the recovery strategy, i.e. $p_c(\gamma = 0)$. Thus, the system would not need a
    strategy to avoid collapsing to the right of this line (region II). The remaining area is the
    range of values of $p = p_A(0)$ and $\gamma$ where the recovery strategy avoids the breakdown 
    and fully restores the system to a functional state.} 
    \label{phase-space}
\end{figure}

In Fig.~\ref{phase-space}, we show the phase diagrams for the interdependent directed networks
that we analyzed earlier (see Fig.~\ref{Pinf}).
Each one of the curves (solid lines) corresponds to a different value of $q$ and represents
the critical value of the recovery probability, $\gamma_c(p)$, such that the system collapses
for $\gamma \leq \gamma_c$ ($P_{\infty} = 0$ in region I). Otherwise, if the strategy is 
sufficiently effective, i.e. $\gamma > \gamma_c$, the system is completely restored 
($P_{\infty} = 1$). In this case, we differentiate two regions for clarity. The curve 
$\gamma_c(p)$ decreases as we increase the value of $p$, until it reaches zero at a given 
point. This point corresponds to the critical threshold in the absence of a strategy, i.e.
$p_c(\gamma = 0)$ such that $P_{\infty}(p_c(\gamma = 0)) = 0$. If we continue increasing 
the value of $p$, the size of the MGSCC at the end of the process becomes different from zero
and starts increasing, regardless of the effectiveness of the recovery strategy. In this 
region of the diagram, the implementation of our strategy is not crucial to avoid the 
collapse of the system. The dashed lines in Fig.~\ref{phase-space} mark the beginning of
this region (region II, where $P_{\infty} > 0$ for $\gamma = 0$ and $P_{\infty} = 1$ for
$\gamma > 0$). Within the remaining area, delimited by the curves $\gamma_c(p)$ and
$p = p_c(\gamma = 0)$, our strategy achieves the most relevant results, avoiding total
breakdown ($P_{\infty} = 0$ for $\gamma = 0$) and recovering the functionality of the system
in its entirety ($P_{\infty} = 1$ for $\gamma > 0$). In addition, by observing the phase
diagrams it becomes more notorious the effect that the fraction of nodes with dependencies in
each network, $q$, has on the outcome of the cascading process and on the effectiveness that
our strategy can achieve. As we commented when analyzing the results from Fig.~\ref{Pinf}, a 
smaller value of $q$ (Fig.~\ref{phase-space} (a) and (c)) hinders the propagation of failures
between networks, because networks become less dependent on each other. However, the region 
in which the strategy is most useful, avoiding system collapse, is significantly reduced 
compared to the fully interdependent case, where $q = 1$ (Fig.~\ref{phase-space} (b) and (d)).

\subsection*{Empirical networks}

\begin{figure}[htb]
    \centering
    \subfloat{\begin{overpic}[width=0.5\linewidth]{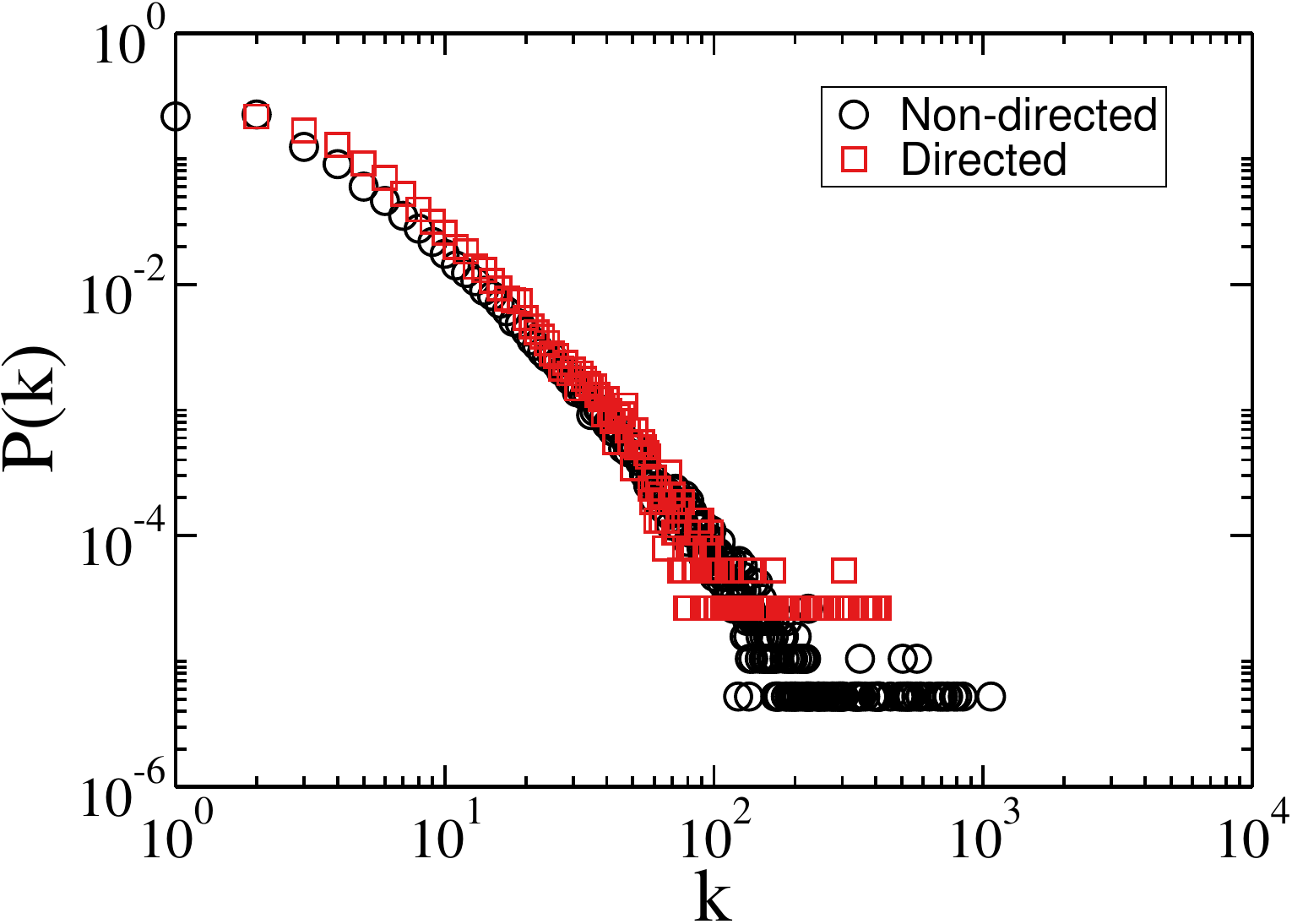}
        \put(0,0){\bf{(a)}}
    \end{overpic}}
    \subfloat{\begin{overpic}[width=0.5\linewidth]{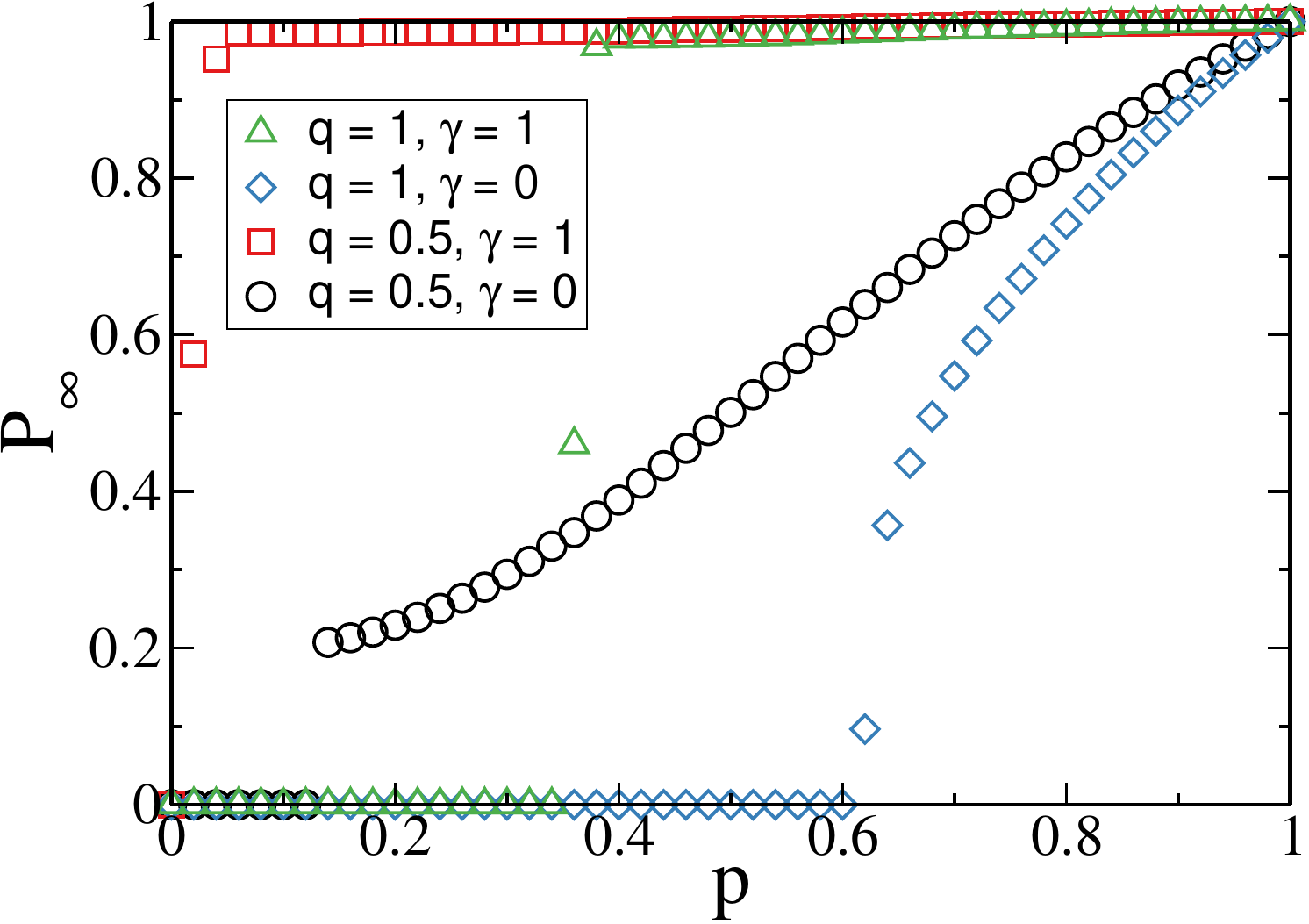} 
        \put(0,0){\bf{(b)}}
    \end{overpic}}
    \caption{Application to networks based on empirical data. (a) Degree distribution of the
    communication network (``tech-RL-caida'' obtained from~\cite{data1}). The original 
    non-directed distribution (circles) has $k_{min} = 1$, $k_{max} = 1071$, and
    $\valmed{k} \simeq 6$, while the processed directed network (squares) has $k_{min} = 2$,
    $k_{max} = 409$, and $\valmed{k} \simeq 7.7$, for both in and out-degree distributions.
    The latter represents a directed network of size $N = 37777$, in which all nodes belong
    to the GSCC. (b) Simulation results for two interdependent networks with the degree
    distribution of in and out-links described in (a). The results correspond to an average
    over 10$^2$ realizations, where interdependencies between networks are randomly assigned.} 
    \label{emp-results}
\end{figure}

In order to demonstrate the applicability of our model, we simulate the process of cascading failures
with recovery of contour nodes in an interdependent system built from an empirical, router-level, 
communication network~\cite{data2,data3} obtained from~\cite{data1}. Since the original network is 
non-directed, we randomly turn the connections into directed links and work only with the emerging 
GSCC. Besides, we remove nodes with $k_{in} \leq 1$ or $k_{out} \leq 1$ in order to begin the process
with a robust GSCC, as we did with the SFc networks from Fig.~\ref{Pinf} (c) and (d). In 
Fig.~\ref{emp-results} (a), we show the degree distribution $P(k)$ before and after modifications.
Then, we use the processed network for both layers A and B of our interdependent system, randomly 
interconnecting nodes from different layers through dependencies and ensuring a fraction $q$ of nodes
with dependencies in each network. In Fig.~\ref{emp-results} (b), we can observe that the results for
the interdependent networks based on empirical data are qualitatively similar to those obtained 
earlier for synthetic networks.

\subsection*{Contrast with random recovery}

\begin{figure}
    \centering
    \subfloat{\begin{overpic}[width=0.5\linewidth]{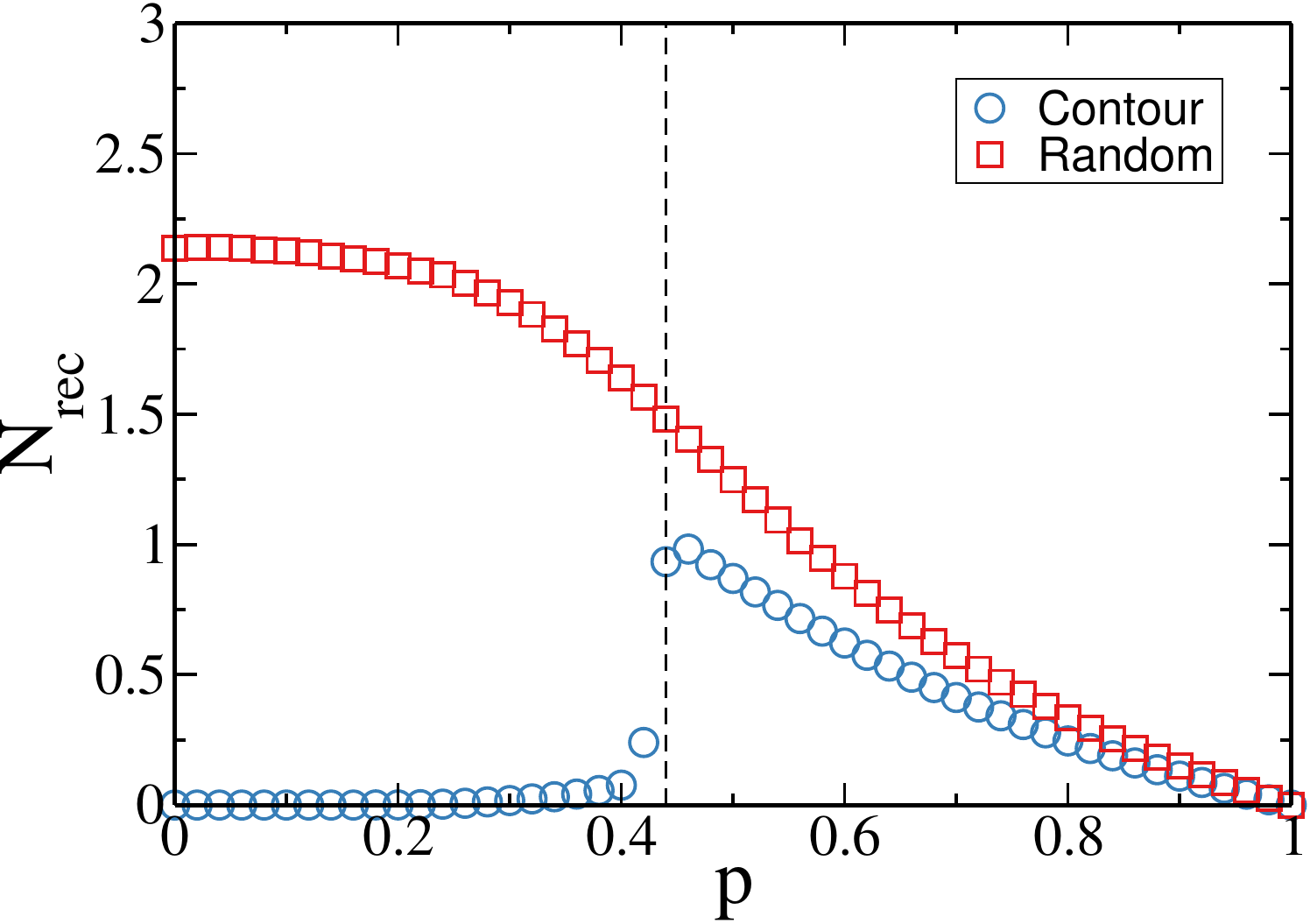}
        \put(0,0){\bf{(a)}}
    \end{overpic}}
    \subfloat{\begin{overpic}[width=0.5\linewidth]{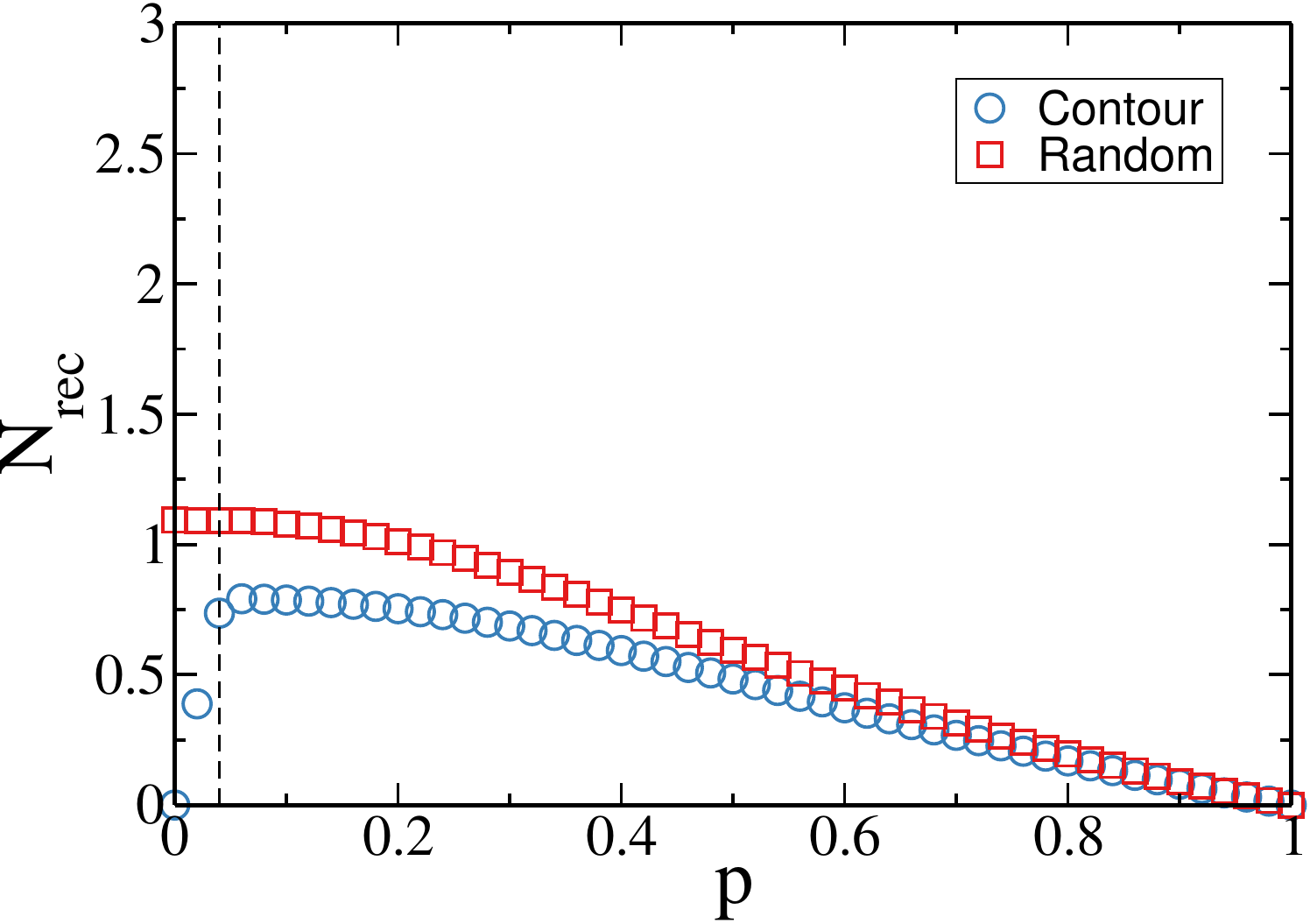} 
        \put(0,0){\bf{(b)}}
    \end{overpic}}
    \caption{Fraction of recovered nodes in the final state, $N_{rec}$, as function of $p = p_A(0)$,
    for (a) $q = 0.5$ and (b) $q = 1$. The fraction $N_{rec}$ is related to the system size $2N$.
    Above the threshold $p^*$ (dashed lines), both strategies completely restore the system, but the
    one that repairs contour nodes requires less amount of resources to do it. The results correspond
    to an average over 10$^2$ realizations on the interdependent system built from empirical data and
    shown in Fig.~\ref{emp-results}.} 
    \label{two-strat}
\end{figure}

Here we present a brief comparison of the proposed recovery strategy with the random repairing
of nodes. For this purpose, we simulate cascading failures in the interdependent system based 
on empirical data (Fig.~\ref{emp-results}) and compute the amount of recovered nodes throughout
the entire process, until the system either collapses or is entirely restored, for both 
strategies. Note that, for the case $\gamma = 1$, the random recovery strategy repairs all 
failures at once, restoring the system immediately to a fully functional state, which may be
quite unrealistic since usually reparations cannot be implemented everywhere at the same time
due to unavailability of resources or technical restrictions. On the other hand, the recovery
of contour nodes is implemented on a restricted set of failures, yielding the previously 
discussed results. Thus, we compare the two strategies in a scenario where the success rate of
recovery is $\gamma = 0.5$ and we show the results in Figs.~\ref{two-strat} (a) and (b), for
$q = 0.5$ and $q = 1$, respectively. We observe that, in both cases, repairing contour nodes
requires less resources when compared to the random recovery strategy. This analysis is 
meaningful for $p > p^*$, with $p^*$ indicated by dashed lines, since below this threshold
the system collapses when the strategy that recovers contour nodes is implemented. Below $p^*$,
the random recovery of nodes can restore the system back to a fully functional state, but in
doing so it consumes an amount of resources equivalent to more than 1.5 times the size of the
entire interdependent system, i.e., $N_{rec} > 3N$, for $q = 0.5$. 
It is worth mentioning that the random recovery of nodes may waste resources in repairing 
failures at a given stage of the cascades that will fail again at the next step due to being
disconnected from the GSCC of their corresponding network. In this way, the proposed strategy
aims to reduce this probability by targeting nodes that, once repaired, become a part of the
GSCC. 

\section{Discussion} 

In this paper, we study the effects of implementing a recovery strategy on a process of 
cascading failures in two interdependent directed networks of the exact same size and degree
distributions. In particular, we analyze the scenario in which the fractions of nodes with
single dependencies (unidirectional or bidirectional) in each network is $q$. The strategy
consists in repairing, with probability $\gamma$, nodes in the contour of the giant strongly
connected component (GSCC) of each layer and is implemented immediately after a cascade of
failures is triggered by the random removal of a fraction $1 - p$ of nodes in one network. We
develop an analytical framework by using node percolation and generating functions, which yield
results that are in well agreement with simulations. We find that, if we sustain the strategy
until there are no more failures or available nodes to repair, the system undergoes an abrupt 
transition between a state of full collapse ($P_{\infty} = 0$) and a state of complete recovery
($P_{\infty} = 1$). For larger values of $q$, it is more likely that the strategy helps to 
avoid collapse in a wider region of the parameter plane $(p, \gamma)$. We also find that more
homogeneous systems are less vulnerable to cascading failures, as the critical threshold $p_c$,
which defines the abrupt transition, is lower compared to heterogeneous interdependent networks 
with the same average degree. Finally, we build phase diagrams in the plane $(p, \gamma)$ by
computing the critical values $\gamma_c(p)$ of the recovery probability needed to recover the 
system. We find three different phases or regions. In the first region, the system collapses 
even though we intervene with our strategy, thus $P_{\infty} = 0$ for all values of $\gamma$. 
The second, most interesting phase, is delimited by the curve $\gamma_c(p)$ and the line
$p = p_c(\gamma = 0)$, where $P_{\infty} = 0$ for $\gamma = 0$ but our strategy achieves the
complete recovery of the system, thus $P_{\infty} = 1$ for $\gamma = 1$. The third region, to 
the right of the line $p = p_c(\gamma = 0)$, comprehends the most robust phase of the system, 
where no strategy is needed in order to avoid full collapse, i.e. $P_{\infty} > 0$ for 
$\gamma = 0$. Our results are qualitatively similar to those of non-directed interdependent 
networks~\cite{dim-15}, although a more detailed comparative analysis would be required in 
order to establish in which kind of systems the proposed strategy is more or less effective.

Moreover, we demonstrate our model of cascading failures with a recovery strategy in an
interdependent system constructed with empirical data of a communication network, and compare
the recovery of contour nodes with a random recovery strategy. We find that recovering 
contour nodes may help in reduce the amount of resources needed in order to restore the 
functionality of the system. Our findings could serve to inform the development of more 
robust recovery strategies in real-world infrastructures such as power grids and 
communication networks. In this way, future research could be focused on including additional
features of real infrastructure networks, such as degree correlations of connectivity and
dependency links, or improving the recovery strategy by searching nodes outside the contour 
of each GSCC that, once repaired, can turn back on entire clusters.

\section{Acknowledgments}

I. A. P. and C. E. L. wish to thank to UNMdP (EXA 1193/24), FONCyT (PICT 1422/19) and CONICET,
Argentina, for financial support. We thank Dr. Lautaro Vassallo for insightful discussions and
valuable comments during the preparation of this paper.

\appendix

\section{Directed network modeling}
\label{app:dir-net}

\begin{figure}
    \centering
    \includegraphics[width=0.25\linewidth]{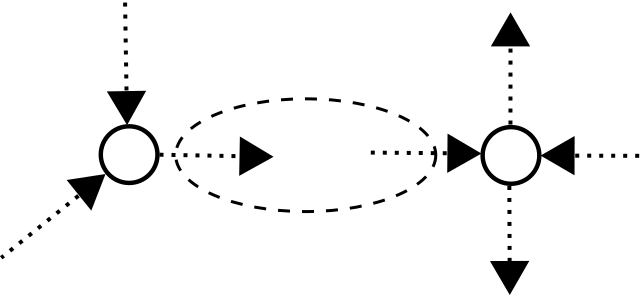}
    \caption{Configuration model for directed networks. The left node has been assigned $k_{in} = 2$
    and $k_{out} = 1$ stubs, while the right nodes has $k_{in} = 2$ and $k_{out} = 2$ stubs, all
    represented by dotted lines. To form a link, the in and out stubs inside the dashed ellipse were 
    picked at random.}
    \label{stubs}
\end{figure}

Considering that an outgoing link, which goes from a source node to a target node, can be viewed as an
incoming link from the point of view of the target node and vice-versa, the condition on the average
degrees of incoming and outgoing distributions for an isolated network, 
$\valmed{k_{in}} = \valmed{k_{out}}$, must be fulfilled. Then, for each one of the $N$ nodes in the
network, we generate random numbers of incoming and outgoing ``stubs'', $k_{in}$ and $k_{out}$, drawn
from the corresponding in and out-degree distributions, since the in-degree is not correlated with the
out-degree. By randomly selecting pairs of in and out-stubs from different nodes, we form directed 
edges (see Fig.~\ref{stubs}) while avoiding bidirectional, multiple, and self-links. As well as in
non-directed networks, this method yields both uncorrelated incoming and outgoing degrees. We can only
successfully match all stubs if $\sum k_{in} = \sum k_{out}$, but for large systems the numbers of in
and out-stubs tend to differ. In order to build the networks in a reasonable amount of time, we set a
maximum tolerance of $10\%$ for this difference.

\section{Finite clusters}
\label{app:fin-clu}

In order to compute the relative size of the finite clusters in both networks at time $t$, 
$FC_A(t)$ and $FC_B(t)$, we follow the evolution of the corresponding GSCC, which is the most
accurate magnitude in our calculations. For instance, in network A prior to the arrival of 
B-failures from time step $t - 1$, the real fraction of functional nodes is 
$\overline{P_{\infty}^A}(t - 1)$, from which we take away the failures produced by nodes in 
the FC\textsubscript{B}, 
$FC_B(t - 1) \left[ q_A q_B + q_A(1 - q_B) \overline{P_{\infty}^A}(t - 1) \right]$, resulting in 
an alternative fraction of remaining nodes in A (different to $p_A(t)$). Then, we obtain 
$FC_A(t)$ by additionally subtracting the fraction of functional nodes in the 
GSCC\textsubscript{A}, 
\begin{equation*}
    FC_A(t) = \overline{P_{\infty}^A}(t - 1) - FC_B(t - 1) \left[ q_A q_B + 
    q_A(1 - q_B) \overline{P_{\infty}^A}(t - 1) \right] - P_{\infty}^A(t).
\end{equation*}
Note that the first term of the failures subtracted from $\overline{P_{\infty}^A}(t - 1)$
corresponds to nodes in FC\textsubscript{B} that have a dependency on and provide support to the
same node from GSCC\textsubscript{A}, the only kind of nodes with which nodes in 
FC\textsubscript{B} can have mutual or bidirectional dependencies, which occurs with probability
$q_A q_B$. The proportion of failures from FC\textsubscript{B} that do not have dependencies but
provide support nodes in GSCC\textsubscript{A} are captured by the second term, 
$q_A(1 - q_B) \overline{P_{\infty}^A}(t - 1)$, which is proportional to 
$\overline{P_{\infty}^A}(t - 1)$ as this kind of failures from FC\textsubscript{B} can also 
provide support already failed A-nodes. The corresponding analysis for the calculation of $FC_B(t)$ is
similar, although it involves finite clusters in A at time $t$ rather than $t - 1$, 
\begin{equation*}
    FC_B(t) = \overline{P_{\infty}^B}(t - 1) - FC_A(t) \left[ q_A q_B + 
    q_B(1 - q_A) \overline{P_{\infty}^B}(t - 1) \right] - P_{\infty}^B(t).
\end{equation*}

%

\end{document}